\documentclass[12pt,preprint]{aastex}
\usepackage{natbib}

\shorttitle{XMM-NEWTON OBSERVATION OF NGC~4649}
\shortauthors{RANDALL, SARAZIN, \& IRWIN}
\slugcomment{Astrophysical Journal, accepted}

\begin{document}

\title{\textit{XMM-Newton} Observation of Diffuse Gas and LMXBs in the
Elliptical Galaxy NGC 4649 (M60)}

\author{Scott W. Randall\altaffilmark{1},
Craig L. Sarazin\altaffilmark{2}, and
Jimmy A. Irwin\altaffilmark{3}}

\altaffiltext{1}{Harvard-Smithsonian Center for Astrophysics, 60
  Garden St., Cambridge, MA, 02138; srandall@cfa.harvard.edu}

\altaffiltext{2}{Department of Astronomy, University of Virginia,
P. O. Box 3818, Charlottesville, VA 22903-0818; sarazin@virginia.edu}

\altaffiltext{3}{Department of Astronomy, University of Michigan,
Ann Arbor, MI 48109-1090; jirwin@astro.lsa.umich.edu}

\begin{abstract}
We present an {\it XMM-Newton} X-ray observation of the X-ray
bright E2 elliptical galaxy NGC~4649.  In addition to bright diffuse
emission, we resolve 158 discrete sources, $\sim$50 of which are
likely to be low-mass X-ray binaries (LMXBs) associated with NGC~4649.
We find evidence for variability in three sources between this
observation and a previous {\it Chandra} observation.  Additionally,
we detect five sources that were not detected with {\it Chandra}
despite its better detection limit, suggesting that these sources have
since brightened.
The total X-ray spectrum of the resolved sources
is well-fit by a hard power-law, while the diffuse spectrum requires a
hard and a soft component, presumably due to the relatively soft diffuse gas
and the harder unresolved sources.
A deprojection of the diffuse emission revealed a radial temperature gradient
that is hot in the center, drops to a minimum at about
20-50\arcsec\ (1.6-4.1 kpc),
and rises again in the outer regions.
The diffuse emission appears to require a two-temperature model with
heavy element abundance ratios that differ from the solar values.
We have verified the existence of
faint radial features extending out from the core of NGC~4649 that had
previously been seen with {\it Chandra}.
The fingers are morphologically similar to radial features seen in
hydrodynamic simulations of cooling flows in
elliptical galaxies, and although their other properties do not match the
predictions of the particular simulations used we conclude that the
radial fingers might be due to convective motions of hot outflowing
gas and cooler inflowing gas.
We also find evidence for a longer, previously undetected filament that
extends to the
northeastern edge of NGC~4649.  The diffuse gas in the region of the
filament appears to have a lower temperature and may also have a
higher abundance as compared to nearby regions.  There also appears
to be an excess of X-ray sources along the filament, though the
excess is not statistically significant.  We conclude that the
filament may be the result of a tidal interaction, possibly with
NGC~4647, though more work is necessary to verify this conclusion.
\end{abstract}


\section{Introduction} \label{sec:intro}

By now it has been well established that the X-ray emission from
early-type galaxies has several components.  These galaxies can
roughly be grouped into two categories based on the ratio of the X-ray
to optical luminosity $L_X/L_B$: the X-ray bright galaxies which have
large values of $L_X/L_B$, and the X-ray faint galaxies which have
small values of $L_X/L_B$.
In the X-ray bright elliptical and S0 galaxies, the X-ray emission is
dominated by thermal emission from interstellar gas
at a temperature of $kT \approx 1$ keV.
The spectra
of their X-ray faint counterparts tend to require two components:
a soft thermal component with $kT \approx 0.3$ keV,
and a hard component which has been fit as either thermal
bremsstrahlung with $kT \ga 5$ keV or a power-law
(Fabbiano, Kim, \& Trinchieri 1994; Matsumoto et al.\ 1997; Allen, di
Matteo, \& Fabian 2000; Blanton, Sarazin, \& Irwin 2001).
The luminosity of the hard component varies roughly
proportional to the optical luminosity, suggesting that the origin of
this component is low-mass X-ray binaries (LMXBs) similar to those
seen in our own Galaxy (Trinchieri \& Fabbiano 1985).  The high
spatial resolution of the modern X-ray observatories {\it Chandra}
and {\it XMM-Newton} has allowed
much of the hard component to be resolved into individual sources,
thereby demonstrating that this hard component is indeed from
individual sources
(e.g., Sarazin, Irwin, \& Bregman 2000, 2001).

In this paper, we present the results of a {\it XMM-Newton} observation
of the X-ray bright elliptical galaxy NGC~4649 (M60).
This is an E2 elliptical galaxy in the Virgo cluster.
NGC~4649 has a close companion galaxy NGC~4647, which
is an Sc galaxy.
This pair of galaxies is also referred to as Arp~116 or VV~206.
With a third more distant galaxy, this pair forms a group of galaxies
WBL~421
(White et al.\ 1999).
NGC~4647 was the host of the Type-I supernova SN1979a
(Barbon et al.\ 1984).

We previously observed NGC~4649 with the ACIS S3 detector on the
{\it Chandra X-ray Observatory}
(Randall, Sarazin, \& Irwin 2004, hereafter Paper I).
We detected 165 discrete X-ray sources with fluxes determined at
$\ge 3 \sigma$ significance, most of which are likely to be low-mass X-ray
binaries (LMXBs) associated with NGC~4649.
The luminosity function of the resolved sources was fit by a broken power-law
with a break luminosity of
$5.3^{+4.4}_{-2.3} \times 10^{38}$ ergs s$^{-1}$.
We argued that the break luminosity might separate black hole binaries (at
higher luminosities) from predominantly neutron star binaries at lower
luminosities.
The cumulative X-ray spectrum of the resolved point sources was fit by a
hard power-law model.
The dominant X-ray emission from NGC~4649 is from diffuse hot gas.
The {\it Chandra} image showed several interesting possible structures
in the X-ray gas, including radial ``fingers'' extending from the center
of the galaxy.
We suggested that these fingers, if confirmed, might be due to convective
motions of the hot interstellar gas, as predicted in some simulations
of elliptical galaxies
(Kritsuk, B\"ohringer, \& M\"uller 1998).
We also detected a 5\arcsec\ (0.41 kpc)
long ``bar'' feature at the center of the
{\it Chandra} image, which we suggested was due to a shock, perhaps driven
by an undetected low-luminosity active galactic nucleus (AGN).

Subsequently, we observed NGC~4649 with {\it XMM-Newton}.
The purposes of this observation included obtaining more detailed
X-ray spectral information on the gas and X-ray binaries, and studying
the structures in the hot ISM.
We adopt a distance for NGC~4649 of 16.8 Mpc,
based on the method of surface brightness fluctuations
(Tonry et al.\ 2001).
This is the same distance assumed for out analysis of the {\it Chandra}
observation, and
is consistent with the corrected recession velocity distance in
Faber et al.\ (1989)
if the Hubble constant is 79 km s$^{-1}$ Mpc$^{-1}$.
Unless otherwise noted, all uncertainties quoted are at the 90\%
confidence level.

\section{Observation and Data Reduction} \label{sec:obs}

NGC~4649 was observed on 2001 January 2 for 53,867 s with the thin
filter in place and with each EPIC instrument in full frame mode.  The
pointing was determined such that the galaxy was centered on the MOS
instruments so that it was not on a node boundary for these
instruments.
The analysis in this paper will be based on data from all three EPIC
instruments: EMOS1, EMOS2, and EPN.  For the MOS instruments, only
events with patterns 0 through 12 (corresponding to single, double,
triple, and quadruple pixel events) were kept, and for the PN only
patterns 0 through 4 (for single and double pixel events) were kept.
We excluded bad pixels, pixels next to bad pixels, bad columns, and
columns next to bad columns or chip node boundaries.
The exposures for the MOS and PN instruments were 51,348 s and 48,951 s
respectively.

Due to their high orbits,
{\it Chandra} and {\it XMM-Newton} experience periods of high background
(``background flares'').
An inspection of the binned light curves from each EPIC instrument showed that
our data were affected by a few strong flares.  To remove them, we first
excluded periods most obviously affected by flares.  The resulting data
were further filtered such that only bins with counts within 2-$\sigma$
of the mean were kept.  This excluded periods of background flares and
data drop-outs.  The remaining exposure times were about 45,800 s and
40,800 s for the MOS and PN instruments respectively.

The PN camera is known to be susceptible to detecting a significant
number of photons during the readout of the chip.  These out-of-time
events were clearly visible in the uncorrected raw image from the PN
camera.  Fortunately, the Science Analysis Software ({\sc sas\footnote{See
\url{http://xmm.vilspa.esa.es/external/xmm\_sw\_cal/sas\_frame.shtml}
.}}) package
provides the script {\sc epchain}, which can be used to produce a
statistical description of out-of-time events.  The resulting
out-of-time image
was scaled by 6\%, which is appropriate for full frame observing
mode\footnote{\url{http://xmm.vilspa.esa.es/external/xmm\_user\_support/documentation/sas\_usg/USG/index.html}},
and subtracted from the raw PN image.  This appeared
to remove all of the obvious artifacts associated with out-of-time events.

Three of the X-ray sources detected have optical counterparts listed
in the U.S. Naval Observatory (USNO) B1.0 optical catalog (Monet et
al.\ 2003).  The X-ray and optical position all agree to better than
3\arcsec.  Thus, we believe that the absolute positions derived from
the X-ray observations are typically accurate to within 3\arcsec.

\section{X-Ray Image} \label{sec:image}

\begin{figure}
\plotone{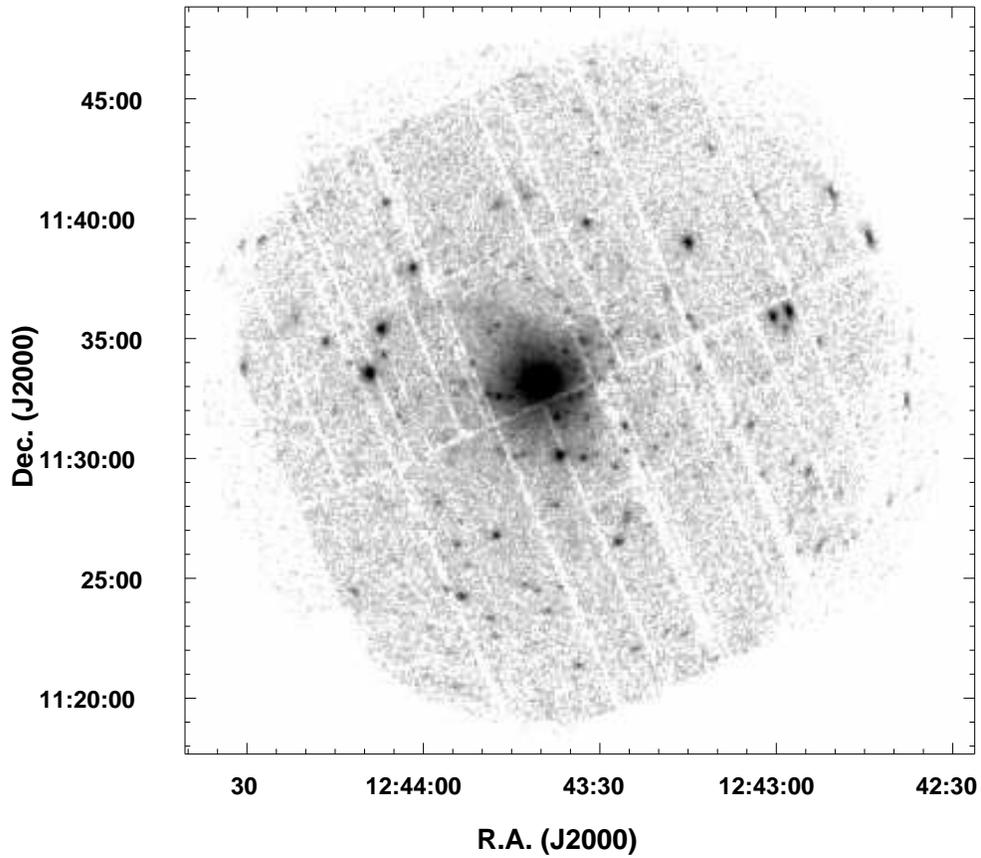}
\caption{Raw {\it XMM-Newton} image of NGC~4649, cleaned of background
flares but uncorrected for background or exposure.  The images from
the MOS1, MOS2, and PN cameras have been overlaid, and the PN image has
been corrected for out-of-time events.
The image was smoothed with a 2 pixel gaussian to make the point sources
easier to see.
The greyscale is logarithmic and ranges from about 0.5 to
30 cnt pix$^{-1}$.
Both discrete sources and diffuse emission are visible.
\label{fig:xray_whole}}
\end{figure}

The raw X-ray image is shown in Figure~\ref{fig:xray_whole} for the
cleaned exposure in the 0.3--12.0 keV band.  The images from the three
EPIC cameras have been superimposed.  The sky background has
not been subtracted from this image, nor has the exposure map been
applied.  This image has been smoothed with a 2-pixel gaussian to
make the point sources more visible.
Many discrete sources are evident.
Also visible is bright diffuse emission at the center of the galaxy,
which swamps emission from discrete sources in that region.
Additionally, there appears to be a faint filament of diffuse emission
extending from the northeast of NGC~4649 and curving to the east.
This feature is discussed further in \S~\ref{sec:filament}.

\begin{figure}
\plotone{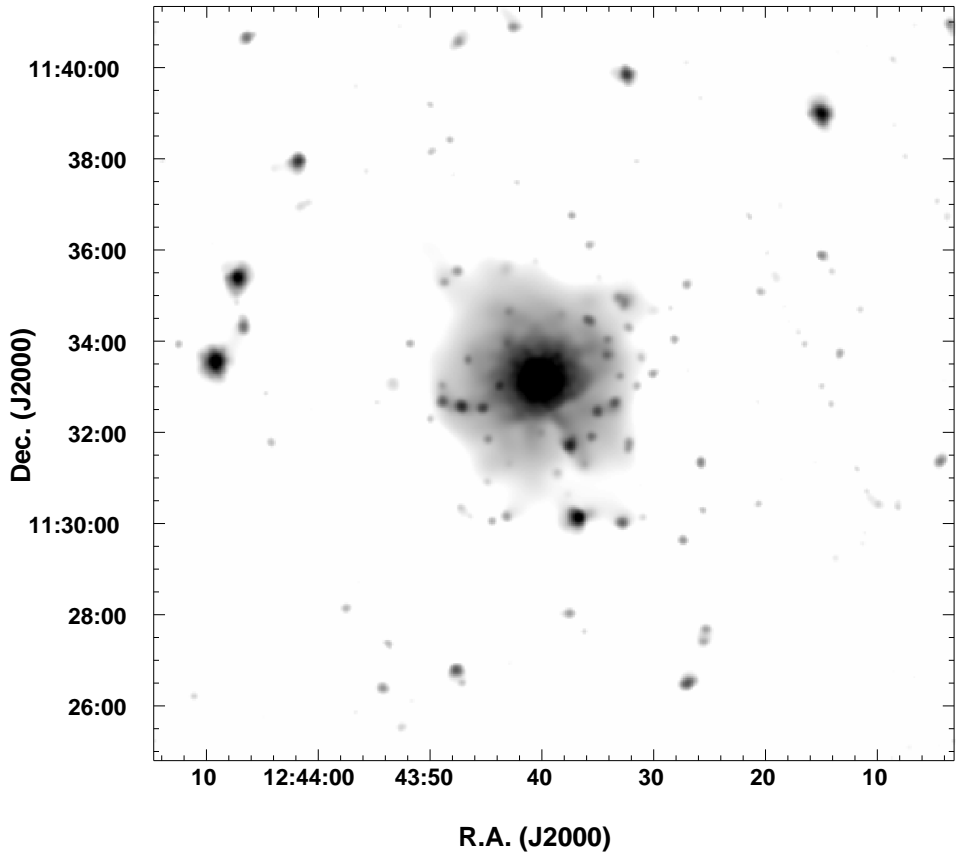}
\caption{Adaptively smoothed {\it XMM-Newton} image of NGC~4649 using
MOS1, MOS2, and PN data,
cleaned of background flares and corrected for exposure and background.
The greyscale is logarithmic and ranges from 2.2$\times 10^{-5}$
to 4.7$\times 10^{-4}$ cnt sec$^{-1}$ pix$^{-1}$.
\label{fig:xray_smo}}
\end{figure}

In order to image the fainter, more diffuse emission, we adaptively
smoothed the central region of Figure~\ref{fig:xray_whole}
to a minimum signal-to-noise ratio of 3 per smoothing beam. The
resulting image is shown in Figure~\ref{fig:xray_smo}.
The image was corrected for exposure and background.
This image shows rather extended diffuse emission from NGC~4649 as
well as the point sources.
In addition to some extended features in the outer parts of the
galaxy,
the image appears to show faint radial ``fingers'' of emission reaching
out from the center of NGC~4649 similar to those seen in the smoothed
{\it Chandra} image, which are discussed further in
\S~\ref{sec:fingers}.

\section{Resolved Sources} \label{sec:sources}

\subsection{Detections} \label{sec:detect}

We opted to use the {\sc ciao}\footnote{See
\url{http://asc.harvard.edu/ciao/}.} wavelet detection program
{\sc wavdetect} to determine the discrete source population.  Sources were
confirmed with a local cell detection algorithm implemented by the
{\sc sas} routine {\sc eboxdetect}.
Source detection was performed in four energy bands: 0.3--1.0 keV,
1.0--2.0 keV, 2.0--12.0 keV, and 0.3--12.0 keV; data from all cameras were used
simultaneously to allow for the detection of faint sources.  The
detection significance threshold was set such that $<$1 false source
(due to statistical fluctuations in the background) would be detected.
  Since {\sc wavedetect} cannot correct
for {\it XMM-Newton's} PSF, the determined count rates were unreliable,
although the detection process itself was unaffected.  We therefore ran
the {\sc sas} routine {\sc emldetect} on the source list generated by
{\sc wavdetect}, which uses a maximum-likelihood algorithm to
determine source extent and count rate (at the time, also allowing the source
positions to vary was not supported) and can properly account for
the PSF.  Thus, the final source fluxes have been corrected for
exposure and the instrumental PSF.
We used the {\sc sas} task {\sc esplinemap} to generate background
maps to be used by {\sc emldetect}.  This task removes specified source
regions from the input image and performs a spline fit to the
remaining data to interpolate across source regions and generate a
background map.
It should be noted that it was
necessary to remove Src.~1 from the list of sources to be cut out of
the background, since this source is not a resolved source but a detection
of the peak in the diffuse emission.  Masking this source when
generating a background map caused the background to be significantly
underestimated for regions just outside the region of the central
source.  Additionally, we found it necessary to increase the minimum
detection likelihood of sources that were removed from the background
map above the typical value for the final source list.  If the
detection likelihood was set too low the background from the bright
diffuse emission was undersampled, which caused the interpolated
background map to underestimate this
component of the background (note that the {\it XMM-Newton} PSF is large
compared to the scale of the radial brightness gradient of the diffuse
galactic emission).
The final source list was trimmed
such that only sources whose fluxes were known to better than
3-$\sigma$ in at least one energy band were kept.

This technique resulted in 158 detected sources.
Blank-sky observations suggest that many of these sources are likely
to be background sources unrelated to NGC~4649 (Hasinger et al.\
2001).  For the purposes of this paper we therefore limit the sources
we consider to those within 328\arcsec\ (27 kpc), or 4 optical effective radii
($R_{\rm eff} = 82\arcsec$, van der Marel 1991) of the center of NGC~4649.
Table~\ref{tab:src} lists the 47 sources contained within this region.
Columns 1-7  give
the source number,
the IAU name,
the source position (J2000),
the projected distance $d$ from the center of NGC~4649,
the 0.3--12.0 keV total count rate
(combined rate for all chips on which source was not on a chip gap or
bad column)
and the 1-$\sigma$ error,
and
the signal-to-noise ratio (SNR) for the total count rate.
Columns 11-13
indicate which of the four energy bands (soft, medium, hard, and
total) each source was detected in for each instrument, where a source is
considered to be ``detected'' if the SNR in that energy band is at
least 3.
All sources were detected in the total 0.3--12.0 keV energy
band using the combined data from each instrument on which the source
was visible (i.e., not on a bad column or chip gap).
The two sources that were not visible on all three EPIC instruments
are marked with a ``i'' in the Notes column of Table~\ref{tab:src}.
Both of these sources were visible on the MOS cameras but not on the PN.
Since we did not detect a point source at the center of the galaxy,
we adopted the central position from 4.86 GHz radio observations of
R.A.\ =\ 12$^{\rm h}$43$^{\rm m}$40\fs02, and
Dec.\ =\ +11\arcdeg33\arcmin10\farcs2 (J2000; Condon, Frayer, \& Broderick 1991),
which is the same value we used for our {\it Chandra} source list for NGC~4649.
The uncertainty of this position is $\la 1\arcsec$.
The statistical errors in the X-ray source positions are typically about
$\pm$1\arcsec\ (1-$\sigma$ errors).

The detection limit in the full 0.3-12.0 keV band was about $1.0\times
10^{-3}$ cnt s$^{-1}$ for $120\arcsec < d < 328\arcsec$.  The
detection limit was increased for $d < 120\arcsec$ (10 kpc) due to the bright
diffuse emission.  This effect is obvious in Table~\ref{tab:src},
where no sources were detected within 66\arcsec\ (5.4 kpc) of the center of
NGC~4649 (except Src.~1, which is not a resolved source but a
detection of the peak in the diffuse emission).

The count rates for the sources were converted into unabsorbed luminosities
(0.3--12 keV) assuming that all of the sources were at the distance of
NGC~4649, which we take to be 16.8 Mpc (Tonry et al.\ 2001).  We
adopted the best-fit {\it XMM-Newton} X-ray spectrum of the resolved
sources within the inner four effective radii (Table~\ref{tab:spectra}
below).  The factor for converting the count rates (0.3--12.0 keV) to
luminosities depended on which instruments contributed to the count
rates (i.e., in which EPIC cameras each source was visible).
Due to the relatively small number (about 30) of sources available
that were both
likely to be associated with NGC~4649 and in a region of relatively
faint diffuse emission we were unable to place useful constraints on
the luminosity function of the sources.

\subsection{Identifications} \label{sec:id}

As noted above, there was only one source, Src.~1, detected within
66\arcsec\ of the center of NGC~4649.
We believe that
this detection does not represent an individual source, but rather a
structural feature in the diffuse emission.
This source is 3.6 times wider (FWHM) than the PSF of a point source
at the same location, and is wider than any other detected source in
the image.
We will therefore drop Src.~1 from further discussion of the point sources.

\begin{figure}
\plotone{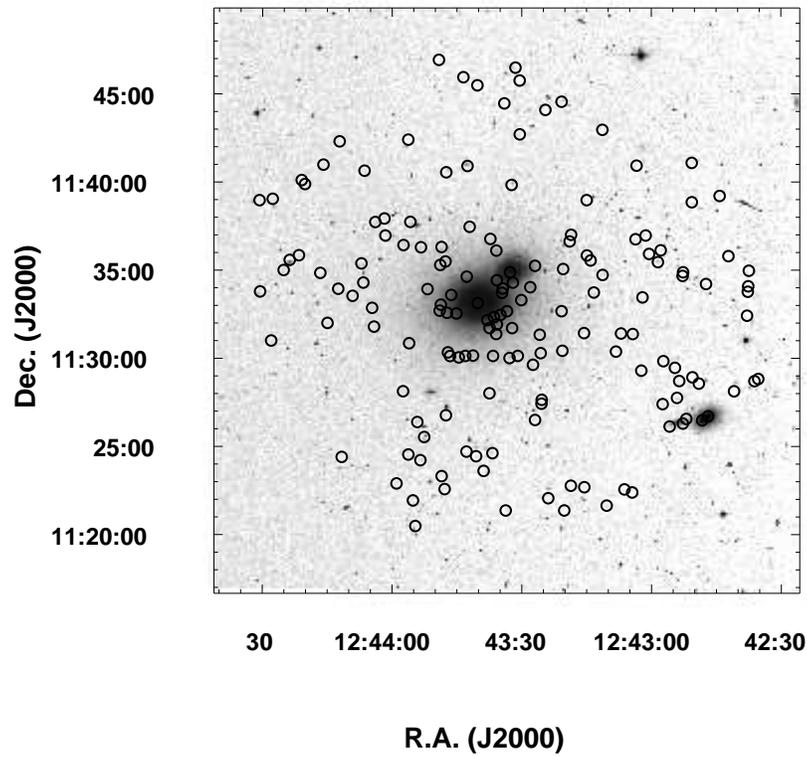}
\caption{DSS optical image of NGC~4649 (M60), and the smaller Sc galaxy
NGC~4647 (northwest of NGC~4649).
The circles indicate the positions of the detected X-ray sources.
\label{fig:dss_field}}
\end{figure}

We compared the positions of the X-ray sources with the
Digital Sky Survey (DSS) image of this region
(Fig.~\ref{fig:dss_field}).
Seven of the sources had possible faint optical counterparts on
this image.
These are all marked with a ``h'' in the Notes column of Table~\ref{tab:src}.
Two of the optical counterparts (for Srcs.~29 \& 41) have positions
listed in the USNO-B1.0 catalog (Monet et al.\ 2003).  Sources with
possible USNO-B1.0 counterparts are
indicated by a ``j'' in the Notes column of Table~\ref{tab:src}.  Two
of the USNO-B1.0 counterparts (for Srcs.~29 \& 47) match the X-ray
positions to within 3\arcsec, while from the density of sources in the
field less than one match is expected
at random (this does not include the match for Src.~21, which is also
within 3\arcsec\ but may be a detection of the nucleus of the
companion galaxy NGC~4647).

We also compared the positions of the X-ray sources to the list of
near-infrared point sources provided by the 2MASS All-Sky Point Source
Catalog
 and
to the list of extended sources in the 2MASS All-Sky Extended Source
Catalog.
Only one source, Src.~21, was within 5\arcsec\ of a 2MASS
source (for the point sources less than 0.5 sources are expected match
at random).
This source, which as mentioned above is within 3\arcsec\ of
the center of NGC~4647, has a match in both the point source and
extended source catalogs.
It is marked with a ``k'' in the Notes column of Table~\ref{tab:src}.

The source list from our {\it Chandra} observation of NGC~4649 was
also correlated with the {\it XMM-Newton} X-ray source list.  There were 23
matches within 5\arcsec, whereas 2 would have been expected at
random.
{\it XMM-Newton} sources with a single {\it Chandra} match are marked
with a ``c'' in the Notes column of Table~\ref{tab:src}.
Some of the {\it XMM-Newton} sources had more than one {\it Chandra} match
and may therefore be a blend of sources.
These sources are marked with a ``d'' in the Notes column.
Three sources had {\it Chandra} counterparts that had previously
been identified with globular clusters (GCs).
Sources corresponding to GCs are marked with a ``e'' in the
Notes column of Table~\ref{tab:src} (note that Src.~12 has two
potential {\it Chandra} source matches, both of which correspond to
GCs).

As shown in Figure~\ref{fig:dss_field}, NGC~4649 has a nearby
companion galaxy, the Sc galaxy NGC~4647.  Some of the X-ray sources
are projected on the optical image of this galaxy and might therefore
be associated with it.  In particular, the position of Src. 21 is
within 3\arcsec\ (0.24 kpc) of the center of NGC~4647 given by optical
observations.  These sources are marked with a ``g'' in the Notes
column of Table~\ref{tab:src}.  This galaxy was also the host of the
Type~I supernova SN~1979a (Barbon et al.\ 1984), although this source
was not detected in our observation.

\subsection{Variability} \label{sec:variable}

We searched for variability in the X-ray emission of the resolved
sources over the duration of the {\it XMM-Newton} observation, excluding
periods with background flares, using the Kolmogoroff-Smirnov test
(see Sarazin, Irwin, \& Bregman 2001). The test was performed on each
camera separately and on the combined data.  For only one of the
sources in Table~\ref{tab:src}, Src. 12, was the probability of being
constant $\ll 1$\%.  This source is marked with a ``f'' in the Notes
column of Table~\ref{tab:src}.  The variability was only significantly
detected when the data from all three chips were used.  This source
was roughly 28\arcsec\ from a chip gap on the PN camera, which is
larger than the
local PSF and the $\pm$0.5\arcsec\ magnitude of the pointing drift
indicated by our optical monitor data.
It varies in brightness by about $\pm$50\% of the average value.
We also compared the luminosities of sources detected both with {\it
XMM-Newton} and with {\it Chandra} (see \S~\ref{sec:id}).  We selected only
those sources that were not a blend of {\it Chandra} sources and were
not near a chip gap or bad column on any EPIC or {\it Chandra}
instrument;
after correcting for the difference in detection bands
we found that 3 out of the 8 selected sources (Srcs.~14, 16, and 28)
had different
luminosities in the {\it Chandra} and {\it XMM-Newton} observations at the
$>$ 3-$\sigma$ level.  In the {\it XMM-Newton} observation Srcs.~14 and 28
brightened, by 52\% and 112\% respectively, while Src.~16 was 63\%
fainter.
However, it should be noted that the uncertainties in the
{\it XMM-Newton} luminosities may be larger than the statistical errors,
due to the difficulties associated with generating a proper
background map for the central region, which are described above in
\S~\ref{sec:detect}.
Additionally, source confusion may artificially inflate some source
luminosities, given {\it XMM-Newton's} relatively large PSF.
Despite the higher detection limit, we detected 5 sources in the
0.3--12.0 keV band that
were not detected in the {\it Chandra} observation.
These were Srcs.~3, 11, 36, 40, and 42.
These sources
have most likely brightened from luminosities below the {\it Chandra}
detection limit ($7\times10^{37}$ ergs s$^{-1}$ in the 0.3--10 keV band).

\subsection{Hardness Ratios} \label{sec:hardness}

We determined X-ray hardness ratios for the sources, using the same
techniques and definitions we used previously
(Sarazin et al.\ 2000;
Paper I).
Hardness ratios or X-ray colors are useful for crudely characterizing the
spectral properties of sources, and can be applied to sources which are
too faint for detailed spectral analysis.
We define two hardness ratios as H21 $\equiv ( M - S ) / ( M + S )$
and H31 $\equiv ( H - S ) / ( H + S ) $, where $S$, $M$, and $H$ are
the
net counts in the soft (0.3--1 keV), medium (1--2 keV), and hard
(2--12 keV) bands, respectively.
The hardness ratios are listed in columns 9 and 10 of
Table~\ref{tab:src} for all of the resolved sources.
The errors in the hardness ratios are determined from the Poisson errors
in the original counts in the bands, and are carefully propagated so as to
avoid mathematically impossible hardness ratios;
that is, the error ranges are limited to $-1$ to 1.
The hardness
ratios were computed using the total net counts from all three EPIC
instruments where possible;
for the two sources not visible on the PN camera (marked with a ``i'' in the
Notes column of Table~\ref{tab:src}) the counts
from the MOS instruments were used.

\begin{figure}
\plotone{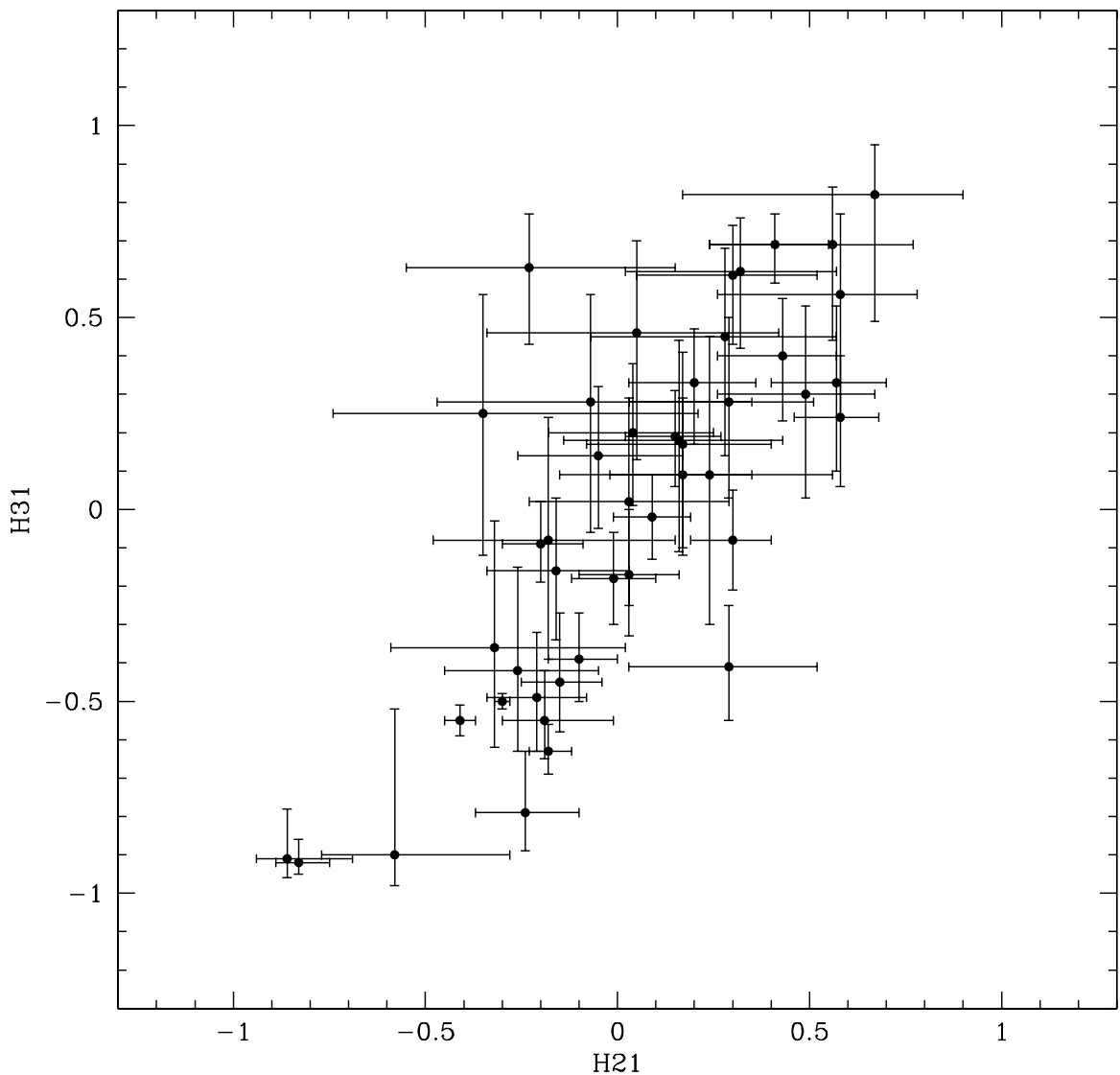}
\caption{Hardness ratios for the NGC~4649 sources visible on all three EPIC
instruments.
Here, $H21 \equiv ( M - S ) / ( M + S )$
and $H31 \equiv ( H - S ) / ( H + S ) $, where $S$, $M$, and $H$ are
the
net
counts in the soft (0.3--1 keV), medium (1--2 keV), and hard
(2--12 keV) bands, respectively.
The 1-$\sigma$ error bars are shown.
\label{fig:colors}}
\end{figure}

Figure~\ref{fig:colors} plots H31
vs.\ H21 for all the discrete sources, excluding sources which were
not detected on all three EPIC instruments.
As was also seen in NGC~4697, NGC~1553, the bulge of NGC~1291, and
a {\it Chandra} observation of NGC~4649
(Sarazin et al., 2000, 2001;
Blanton et al., 2001;
Irwin et al., 2001; Paper I),
most of the sources lie along a broad diagonal swath extending roughly from
(H21,H31) $\approx (-0.5,-0.7)$ to (0.3,0.2).
We find two ``supersoft'' sources ([H21,H31] $\approx$ $[-1,-1]$),
Srcs.~2 and 21.  Src.~2 is located in a region of bright diffuse
emission and therefore, as discussed in \S~\ref{sec:detect}, the
luminosity, and hence the X-ray colors, may be inaccurate.  Src.~21
may be a detection of diffuse gas in the companion galaxy NGC~4647, which
would explain the soft colors.
One source, Src.~18, had relatively
hard X-ray colors ([H21,H31] $>$ $[0.6,0.5]$), which suggests that
this source may be an unrelated, strongly absorbed AGN.  This
conclusion agrees with our previous results from {\it Chandra}
(this source corresponds to {\it Chandra} Src.~110).

\section{X-ray Spectra} \label{sec:spectra}

We used the {\sc sas}\footnotemark[1]
task {\sc evselect} to extract spectra.  For extended sources the
response matrices were calculated using the {\sc sas} tasks {\sc
rmfgen} and {\sc arfgen}.  For the individual resolved sources local
backgrounds were used.
For the diffuse spectra we used the blank sky background files
E1\_ft0000\_M1.fits, E1\_00ft00\_M2.fits, and E1\_0000ft\_PN.fits
provided by Read \& Ponman (2003).  This is preferable to using
background data extracted from off axis regions of our own observation
since there are significant spatial variations in the background,
particularly on the PN chip
(Lumb 2002\footnote{Available at \url{http://xmm.vilspa.esa.es/docs/documents/CAL-TN-0016-2-0.ps.gz}.};
Read \& Ponman 2003).  The blank sky data were
cast into sky coordinates and then
filtered for flares using the same method described in
\S~\ref{sec:obs}.  Since the particle background may vary from
observation to observation we normalized the blank sky data so that
the count rates in the 10-12 keV band for the MOS cameras, and in
the 12-14 keV band for the PN camera, matched our observed count
rates.  However, this turned out to be a small correction ($< 5\%$).
The PN
spectra were corrected throughout for out-of-time events by
subtracting spectra extracted from the out-of-time images mentioned in
\S~\ref{sec:obs} and scaled by 6\%.
Fits to the source spectrum were done for the full 0.3--12.0 keV band.
Spectral fits for the diffuse emission were restricted to the 0.3-5.0 keV band
since emission from the diffuse gas
(which has $kT \approx 1$ keV) is weak above
$\sim$5 keV.
Also, there are some instrumental lines above 5 keV,
particularly for the PN camera, that apparently do not necessarily
scale with the background,
which can significantly affect the
fitting process (Kirsch 2003\footnote{Available at
\url{http://xmm.vilspa.esa.es/docs/documents/CAL-TN-0018-2-1.ps.gz}.}).
Each spectrum has been grouped to a
minimum of 20
counts per pulse invariant (PI) channel so that $\chi^2$ statistics apply.
We used {\sc xspec} to fit models to the data.
The absorption column was fixed at the
Galactic value ($N_H = 2.20 \times 10^{20}$ cm$^{-2}$; Dickey \&
Lockman 1990) throughout, unless otherwise indicated.

We expect that there may be excess background emission in this region
as was seen in {\it Chandra} observations of NGC~4649.  This excess
may be a combination of emission from the Virgo cluster, the north
Polar Spur, excess particle background, and/or unresolved
background sources (although we expect the blank sky background
correction to largely account for any excess particle background
and for unresolved background sources).  To test for this excess we
examined a region far from the center of NGC~4649 with sources removed
from the region, using the blank sky data as background.
We determined that there was in fact a detectable excess background that
could be fit by an absorbed MEKAL plus power-law model
with temperature $kT = 0.170^{+0.013}_{-0.013}$ keV
(the abundance was essentially undetermined)
and photon index
$\Gamma = 2.13^{+0.07}_{-0.07}$.
Allowing the absorption to vary from Galactic did not significantly
alter the fit.
This model was scaled by the solid angle of the source, frozen, and
included in all spectral fits for diffuse
sources, although in the brighter inner regions of NGC~4649 this
component was negligible.

A summary of the best fitting spectral models
is given in Table~\ref{tab:spectra} for different components of the
X-ray emission.
The first column gives the origin of the spectrum, the second column
lists the spectral model used,
the third column gives the absorbing column,
the fourth and fifth columns give the
temperature $T_s$ and abundances (if relevant) for the softer component
of the spectrum, the sixth column gives the power-law photon spectral
index $\Gamma$ or temperature $T_h$ of the harder component in the
spectrum, the seventh column gives the value of $\chi^2$ and the number of
degrees of freedom (dof), and the last column gives the number of net
counts (after background subtraction) in the spectrum.

\subsection{X-ray Spectrum of Resolved Sources} \label{sec:src_spec}

\begin{figure}
\centering
\includegraphics[angle=-90,width=4.0in]{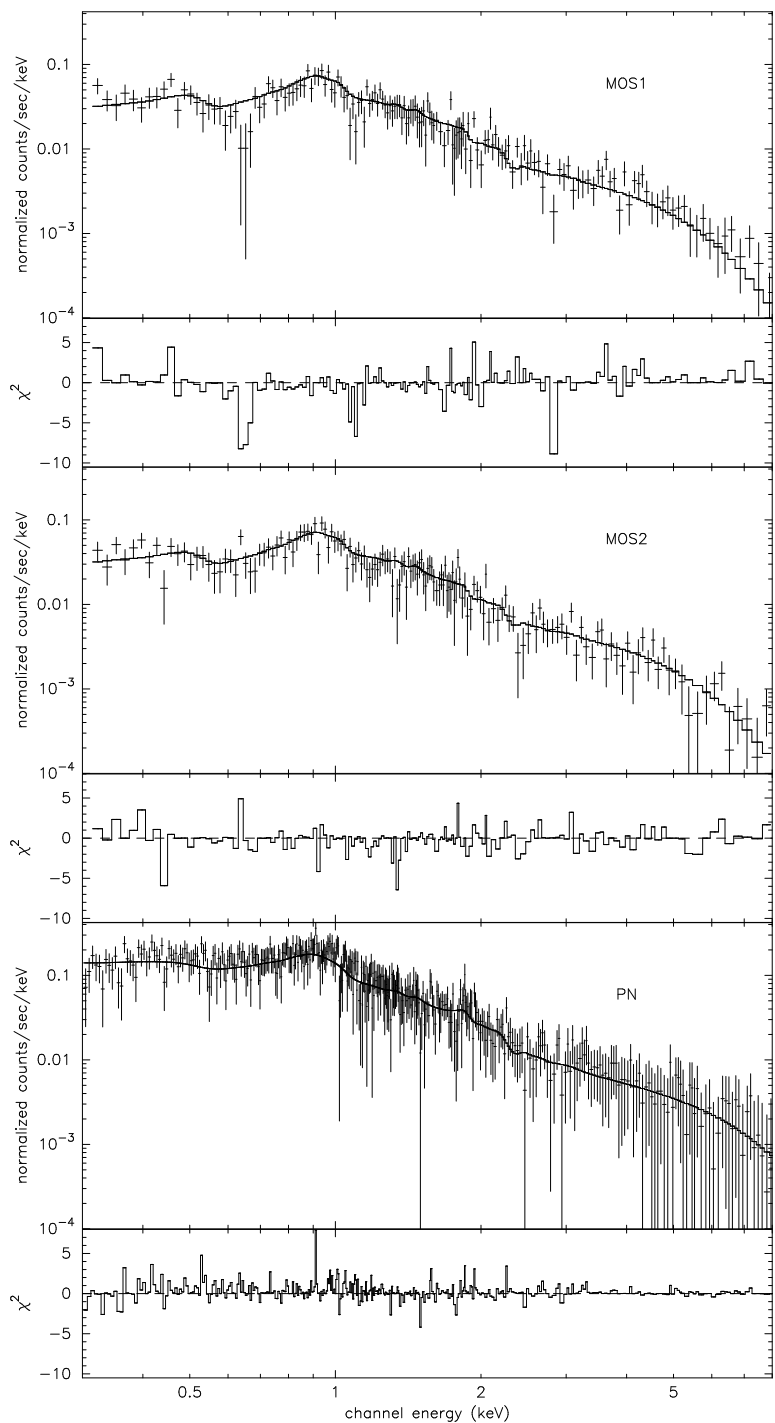}
\caption{
X-ray spectrum of the sum of the sources within four $R_{\rm eff}$ of
NGC~4649, excluding Src.~1, fit with a model combining Galactic
absorption and a hard power-law.  The points with the error bars are
the data and the histogram shows the fitted model.
The lower panel in each plot shows the individual bin contributions
to the chi-squared of the fit.
The top panel shows the fit for the MOS1 camera, the middle MOS2, and the
bottom PN.
\label{fig:src_spec}}
\end{figure}

Figure~\ref{fig:src_spec} shows the spectra of the sum of the sources
within four effective radii ($R_{\rm eff} = 82\arcsec$, 6.7 kpc) of the center
of NGC~4649.  Given the detection limit and density of background
sources (Hasinger et al.\ 2001) we expect to find about 2 background
AGN in this region.
The spectra showed some evidence for weak emission lines, possibly
indicating the presence of hot diffuse gas,
even though local background spectra had been subtracted from the
source spectra.  In general, including a MEKAL or VMEKAL component in
our models fit to the source spectra significantly improved the quality of
the fits, and the best-fit parameter values closely matched those
found for the diffuse gas in \S~\ref{sec:diffuse_spec}.  We conclude
that the local background subtraction did not completely account for
the diffuse emission component, most likely due to {\it XMM-Newton's}
relatively large PSF in combination with the steep brightness profile
seen in the diffuse emission.
Although symmetric local background regions should remove any linear
gradient in the background, they will underestimate the background
if the surface brightness has a negative second derivative (i.e., is concave
downward),  as is true for a beta-model galaxy surface brightness.
Additionally, the large source regions and density of central sources
made it difficult to simultaneously find ideal local background
regions for all of the sources.
On average, these effects led to an under-subtraction of the background
emission.
To compensate, we included the best-fit VMEKAL model to the diffuse
emission within 3$R_{\rm eff}$ (see \S~\ref{sec:diffuse_spec} and
Table~\ref{tab:spectra}) to our models fit to the source spectra.  In
every case we tried including this component improved the quality of
the fit.  The results of model fits are given in Table~\ref{tab:spectra}.

We first considered models in which the spectra of the sources was
represented by a single, hard component.
Initially, we tried a thermal bremsstrahlung model with a temperature
$T_h$, which provided
an acceptable fit
with a $\chi^2$ per
dof of $\chi^2_\nu = 0.85$.  Since our analysis of {\it Chandra} data
somewhat preferred a power-law model for the hard component over a
bremsstrahlung model we also tried a power-law fit.  This provided an
even better fit, with a $\chi^2_\nu = 0.75$.  The best-fit photon
index of $\Gamma = 1.76$ was was very close to that found from the
{\it Chandra} data and for the unresolved sources in
\S~\ref{sec:diffuse_spec}.  This fitted model is shown in
Figure~\ref{fig:src_spec}.  We will adopt this as our best-fit model
for the spectrum of the source population.

\subsection{Diffuse X-ray Spectrum} \label{sec:diffuse_spec}

\begin{figure}
\centering
\includegraphics[angle=-90,width=4.5in]{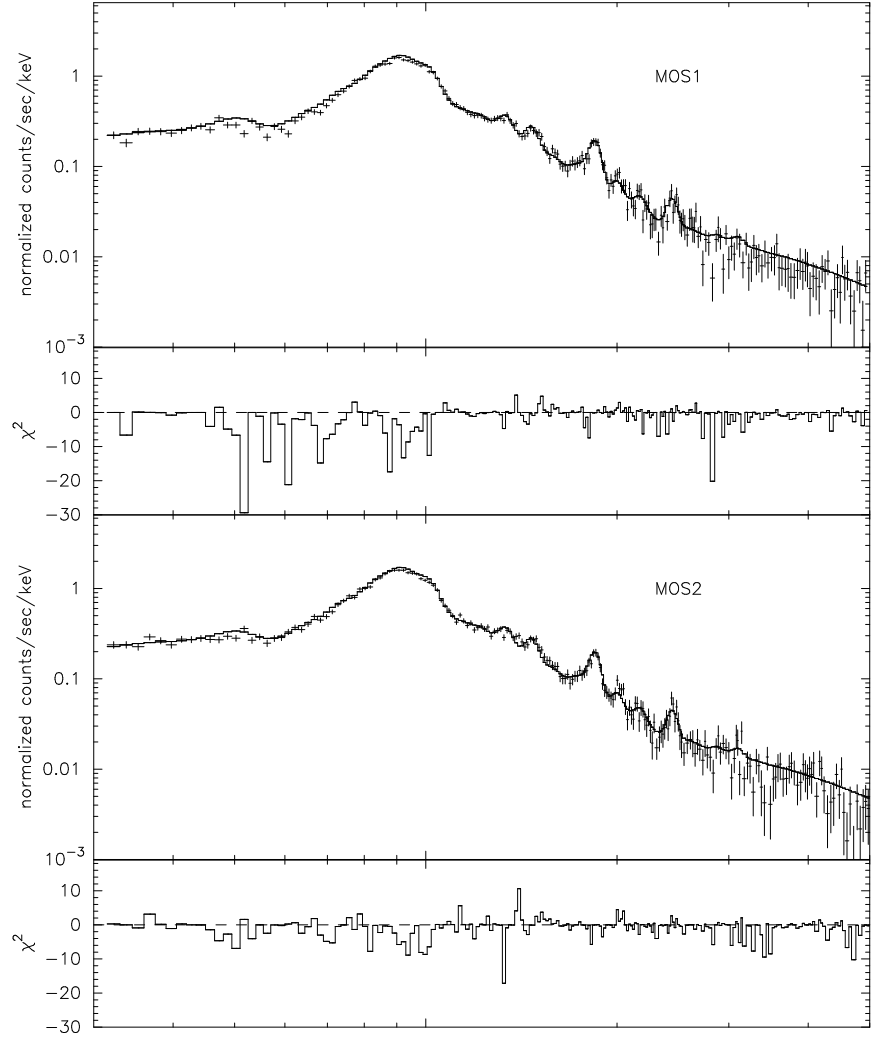}
\caption{X-ray spectrum of the diffuse emission within three $R_{\rm eff}$ of
NGC~4649, fit with a model combining Galactic absorption, a soft
VMEKAL component, and a hard power-law component with photon index
frozen at 1.78.  The
top panel shows the fit for the MOS1 camera and the bottom the MOS2.
\label{fig:diffuse_spec}}
\end{figure}

We extracted the spectra for the diffuse emission from within three
effective radii ($R_{\rm eff} = 82\arcsec$, 6.7 kpc) of the center of NGC~4649
(Figure~\ref{fig:diffuse_spec}), excluding all sources except Src.~1.
Strong emission lines are clearly present in the spectra that
indicate that the emission is mainly thermal emission from hot
interstellar gas, which is consistent with our previous results from {\it
Chandra} (Paper I).
The initial fits to the diffuse emission simultaneously using data
from each of the
EPIC instruments gave large abundance values,
as compared with {\it Chandra}, with very large errors.  Further
analysis showed that the results for the PN data gave larger abundance
values as compared to the MOS data, which were more consistent with
{\it Chandra} results.  Since we were unable to resolve these
inconsistencies we omitted the PN data from all fits to the diffuse
emission.  Removing the PN from fits to the source spectra did not
significantly alter the results, presumably due to the lack of strong
lines in these spectra.

Based on the line emitting spectra we first tried a single
temperature MEKAL model.  This did not provide an acceptable fit.  In
particular, there was a deficit of hard emission in the model for
photon energies $ \ga 1.5$ keV.
Since hard emission is expected from unresolved
sources we added a power-law component to this model, which greatly
improved the quality of the fit (see Table~\ref{tab:spectra}).  The
best fit photon index was $\Gamma = 1.78$, which agrees extremely well
with the value found for the source spectral fits from {\it Chandra} data.
For simplicity the photon index was set equal to this
value and frozen for all subsequent fits to the diffuse emission.

As Table~\ref{tab:spectra} shows, the basic absorbed MEKAL plus
power-law model still does not provide a statistically acceptable fit to the
data.
Since previous work has found evidence for multiple temperature components
and temperature gradients in this and
other elliptical galaxies (e.g., Buote 2002; Paper I),
we added a second MEKAL component to the previous model with the
temperatures and normalizations allowed to vary and
with the abundances of the two MEKAL components tied together.
This provided a significantly improved fit, although it still was
not very good.
We also tried allowing the absorbing column to vary, but this did
not improve the fit significantly.

The two temperature model still did not provide a statistically good
fit to the data (chi-squared per degree of freedom $\chi^2_{\nu} =
2.50$).  An examination of the residuals showed that much of the
contribution to the chi-squared of the fit was coming from regions
near emission lines.
This suggested that the elemental abundances did not simply scale with solar,
confirming the result of our {\it Chandra} analysis.
We therefore tried fitting the data with an absorbed VMEKAL
plus power-law model.  Initially we allowed each abundance value to
vary independently, save that of He which was set to solar.  Although most of the
resulting abundances were very poorly constrained, their values
suggested that the abundances could be divided into three groups: C,
N, O, and Ca; Ne through Ar and Ni; and the third group being Fe by
itself.  We tried various other groupings, including the best grouping
found with the {\it Chandra} data, but none of them provided as good a
fit to the data.  In particular, it seems necessary to allow iron to
vary independently; grouping it with other elements in general worsened the
fit.

Although the grouped VMEKAL plus power-law model provided an
improved fit to the data (see Table~\ref{tab:spectra}) it still did
not give a statistically good fit.
One possible cause of this is the large number of counts in the spectrum
and relatively small statistical uncertainties.
If the systematic errors are comparable to or larger than the statistical
ones, then the $\chi^2$ of the fit may be increased.
Including a systematic error of
5\%,
which is roughly what is expected based on calibration analyses of the
EPIC cameras
(Kirsch 2003),
decreased $\chi^2_{\nu}$ from
1.36 to
1.03 while the best fit
parameter values remained essentially unchanged.
We believe that the systematic errors in the spectral calibration may well
be on the order of 5\% or larger, and thus can affect the $\chi^2$ values.
On the other hand, they do not appear to have an significant effect on the
spectral fit parameters.

The spectra shown in Figure~\ref{fig:diffuse_spec} show particularly
strong lines from S, Si, and Fe.  We therefore determined the Si/Fe
and S/Fe ratios by fitting a modified VMEKAL model to the spectra.
All other elemental abundances were allowed to vary freely, save that
of He which was set to solar.  The Si/Fe ratio is an indicator of what
type of supernova (SN) explosions have enriched the diffuse gas.
We find Si/Fe = 2.38$^{+0.16}_{-0.15}$ and S/Fe = 2.34$^{+0.54}_{-0.19}$.
Model calculations have found that for Type Ia SN the Si/Fe ratio is
$\la1$ (Nomoto et al.\ 1984), while observational studies of metal-poor
Galactic stars find that the average products of Type II SN have
$\alpha$-element abundances 2-3 times larger than the Fe abundance
(e.g., Edvardsson et al.\ 1993; Nissen et al.\ 1994; Thielemann et
al.\ 1996).  We therefore conclude that the diffuse gas in NGC~4649 was
most likely enriched by more by SN~II than by SN~Ia.

\subsection{Radial Variation in the Spectrum of the Diffuse Component} \label{sec:radial}

The spectral properties of the diffuse emission as a function of
radius were examined to
search for radial temperature and abundance gradients in the gas. We
divided the diffuse emission into concentric annuli centered on the
galactic center such that each
annulus had at least 3000 net counts per instrument and was at least
10\arcsec\ (0.81 kpc) wide, and extracted spectra from each annulus (with source
regions subtracted).
The annuli were required to be at least the 10\arcsec\ in width to avoid
effects of {\it XMM-Newton's} PSF, which set the widths of the inner
annuli.
The resulting 9 annuli are defined in Table~\ref{tab:proj_spec_grad}
(emission from larger radii was dominated by the excess background
emission discussed in \S~\ref{sec:spectra}).  For each model discussed
in this section the photon index of the unresolved sources was fixed
at $\Gamma = 1.78$.

\subsubsection{Projected Spectra} \label{sec:proj}

We first attempted to fit the observed spectra uncorrected for projection
effects.
We chose a simple MEKAL plus power-law
model to simplify the comparison with the results from the deprojected
spectra given in \S~\ref{sec:deproj}.
The results are given in Table~\ref{tab:proj_spec_grad}.
The resulting temperature profile is consistent with our results from
{\it Chandra} observations;
the gas temperature is a bit hotter within the central 10\arcsec,
drops to a minimum between 20-50\arcsec\ (1.6-4.1 kpc), and rises again in the outer
regions.
Unfortunately, the abundance values are much too poorly
constrained to allow detection of any radial variations, although one
might note that the abundance of the outer annulus appears to be
lower than that of any other.
The large abundance errors are most likely due to the fact that much
of the emission is in lines so that the continuum emission from H
is faint, leading to large errors in the abundance ratios relative to
hydrogen.
In general,
the normalization factors for the mekal components are poorly
constrained, which one would expect if this were the case.

Table~\ref{tab:proj_spec_grad} also shows that, although some of the
fits at larger radii are adequate, the fits to the central regions are
not.  One possible explanation for this is that when observing the
central regions we must look through the outer regions, which will
in general contaminate the spectra.  Fortunately, {\it XMM-Newton's} large
collecting area provided us with enough counts ($>$3000 per instrument
per annulus) to do a spectral deprojection to account for this effect.
In the next section we describe the deprojection and give the results.

\subsubsection{Deprojected Spectra} \label{sec:deproj}

We first fit the emission in the outermost annulus using the same
MEKAL plus power-law model with Galactic absorption as used for the
projected annular fits.  The best-fit values were then frozen and
added to the fit for the next inner annulus, with the normalizations
adjusted to account for geometric effects.
The model component normalizations were fit and frozen independently for each
camera since the different chip geometries lead to different fractional
area covered by chip gaps for each camera in each annulus.  We
continued this process, fitting annuli and freezing the best fit values
for the fit to the next annulus in, all the way to the center of the
diffuse emission.  The results are given in Table~\ref{tab:deproj_spec_grad}.
The temperature profile is
essentially the same as seen for the projected fits, with a slightly
larger difference in temperatures between the hotter core and the
cooler region just outside the core (presumably because the deprojection has
removed the influence from the gas in outer annuli).  The
$\chi^2_{\nu}$'s of the fits to the central three annuli are
essentially unchanged.
We conclude that while projection
effects can noticeably influence our fitting results in the central
region of NGC~4649, it is unlikely
that they are solely responsible for the poor quality of the fits in
this region.

As indicated in Table~\ref{tab:deproj_spec_grad}, the abundance values
for the fourth and fifth annuli from the center were essentially
undetermined.  It is interesting to note that the projected {\it
  Chandra} spectra show anomalously large abundance errors in the same
region (between 38-54\arcsec, or 3.1-4.4 kpc).  It is possible that
the spectra in this region are contaminated by some unresolved source
that biases the abundance values.  Since the abundances were essential
unconstrained, the best-fit temperatures were only
very slightly affected by varying the abundances over a wide range of
values.  Although it would be interesting to look at this region in
more detail, for the purposes of this paper we simply conclude that
the abundance values are undetermined and that the observed
temperature gradient is not greatly affected by this uncertainty.

We next tried different spectral models for the innermost annulus.
The best-fit models for the total spectrum of the diffuse emission
(\S~\ref{sec:diffuse_spec}) include several temperature components with a
wider range of temperatures than seen in the deprojected annular spectra.
This suggests that the gas may be locally multiphase in its thermal
structure.
Thus, we first tried fitting the deprojected spectra of the innermost
annulus with two MEKAL
components with their abundances set equal.
This provided only a slightly improved but still not ideal fit
($\chi^2_{\nu}$ = 2.06).
As was also  the case with the projected spectra, we found that much of
the contribution to the $\chi^2_{\nu}$ of the fit originated from
regions near lines, suggesting that the individual elemental
abundances do not simply scale with solar.  We therefore tried a
VMEKAL model with the elements grouped in the same way as they were
for the fits to the total diffuse emission
(\S~\ref{sec:diffuse_spec}).
Although this provided only an improved fit, with
$\chi^2_{\nu}$ = 1.45, it is still not a statistically good fit.
We therefore tried a two-temperature VMEKAL model with tied
abundances, thereby allowing for a two-phase ISM with abundance values that do
not scale with solar.  The fit was somewhat improved, with
$\chi^2_{\nu}$ = 1.29.
The best fit temperatures were 1.425$^{+0.321}_{-0.213}$ and
0.741$^{+0.023}_{-0.022}$, while the best fit abundances were
0.0$^{+0.1}_{-0.0}$ for C, N, O, and Ca;
2.7$^{+2.9}_{-1.1}$ for Ne, Mg, Si, S, Ar, Al, Na, and Ni;
and
1.4$^{+1.3}_{-0.3}$ for Fe.
Though this is still not an ideal fit in the statistical sense, we
note that the fit statistic is only somewhat larger than that of the
outermost annulus, where the projection effects are unimportant.
Furthermore, one might expect the goodness of the fit to be degraded
since the models used for the innermost annuli to deproject the
emission from the central region were statistically poor fits to the data.
Lastly, we tried adding a cooling flow component (mkcflow) to our
models in lieu of a second MEKAL or VMEKAL component to model the
cooler emission.  Adding this component only slightly improved the fit
($\chi^2_{\nu} = 1.96$) over the single temperature model, and the fit
was inferior to that of the two-temperature VMEKAL model.
We therefore conclude that the central emission is best modeled by a
two-temperature thermally emitting gas with non-solar abundance ratios.

\subsection{Density Profile} \label{sec:dens}

\begin{figure}
\plotone{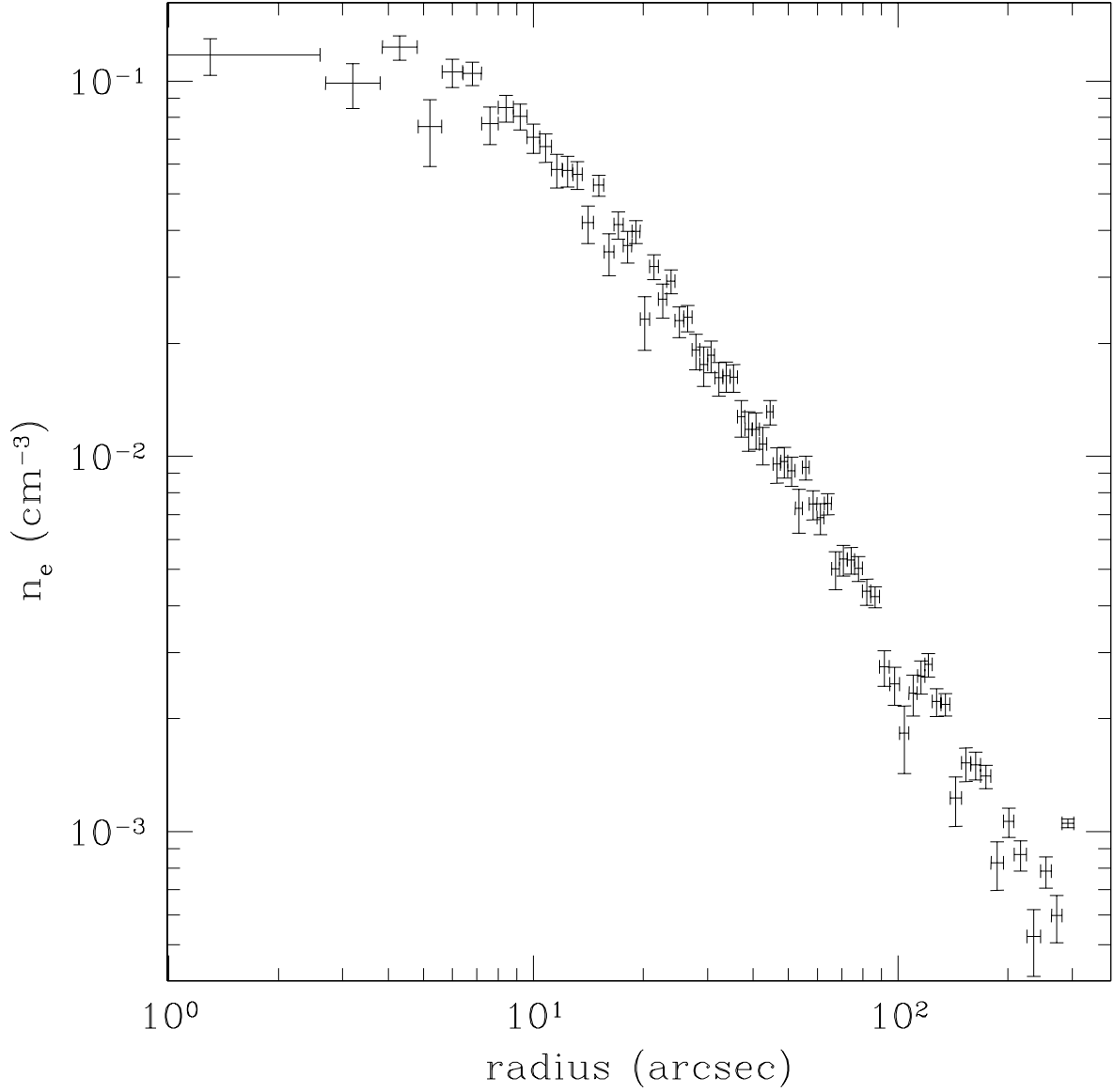}
\caption{
Derived electron number density profile for the diffuse gas, assuming that the
distribution is spherically symmetric and that the gas temperature and
abundance are constant.
\label{fig:density}}
\end{figure}

The X-ray surface brightness profile was determined by accumulating
counts in circular annuli, excluding point sources, and correcting
for both exposure and the blank sky background.
We used 70 annular bins between 0-303\arcsec defined such that each
bin had $\ga 500$ net counts from each MOS camera.
This profile was deprojected to give the gas density profile, assuming
that the temperature and abundance of the gas was uniform.
The emissivity of the gas was derived from the MEKAL component of the
best-fit MEKAL plus power-law spectral model for the total emission
within 303\arcsec.
The resulting electron number density profile is plotted in
Figure~\ref{fig:density}.  It is in reasonable agreement with the
density profile derived by Trinchieri, Fabbiano, \& Kim (1997) using
{\it ROSAT-PSPC} observations.

This procedure assumes that the spectrum of the gas (i.e., the
abundance and temperature) is the same everywhere.  However, the density is
only weakly dependent on the spectrum, and as
Tables~\ref{tab:proj_spec_grad}~\&~\ref{tab:deproj_spec_grad} show
the temperature and abundance do not vary greatly as a function of
radius (although some abundance values are poorly constrained).
Therefore our results should not be strongly affected by this assumption.

\section{Structure in the Diffuse Emission} \label{sec:structure}

In our paper on a {\it Chandra} observation of NGC~4649 we reported
finding evidence for structure in the diffuse emission.  Specifically,
we found faint radial ``fingers'' reaching out from the center of
NGC~4649, and a central X-ray bar, about 5\arcsec\ (0.41 kpc) long, which
appeared to coincide with 4.86 GHz radio emission.  In this section we
present additional results on these features provided by the {\it XMM-Newton}
data.
Additionally, we present evidence for a previously undetected
feature, a faint filament of diffuse emission extending to the
northeast from the center of NGC~4649.

\subsection{Radial Features} \label{sec:fingers}

\begin{figure}
\plotone{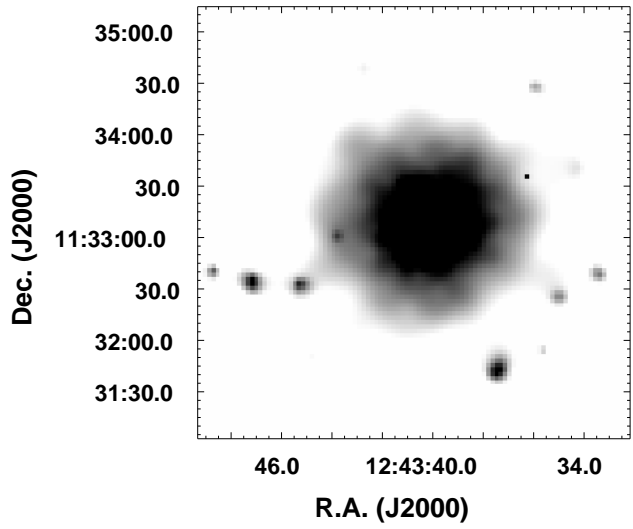}
\caption{
Adaptively smoothed image of the central region
of NGC~4649 created using the MOS1 and MOS2 data,
cleaned of background flares and
corrected for background and exposure.
Faint radial features can be seen extending out from the center of the
galaxy.
The greyscale is logarithmic and ranges from
4.8 to 19.2
cnt pix$^{-1}$.\label{fig:fingers}}
\end{figure}

The smoothed image of the core of NGC~4649 in Figure~\ref{fig:fingers}
shows faint radial fingers of diffuse emission extending from the
core.  Qualitatively, these features appear similar to those seen in
the corresponding {\it Chandra} image, although they are somewhat less
defined due to {\it XMM-Newton's} relatively large PSF.  This
similarity reinforces our previous conclusion that these structures
represent real features, and are not an artifact of the telescope
support structure.  To check the statistical
significance of these features, we defined a set of regions containing
the most obvious fingers, as was done with {\it Chandra}.  The regions
were slightly further from the
center of NGC~4649 than the {\it Chandra} regions to minimize the
impact of blurring the fingers by {\it XMM-Newton's} PSF.  We
considered only the raw data with {\it Chandra} source regions
removed, unsmoothed
and uncorrected for exposure,
to ensure that the fingers were not an artifact of the smoothing
process.  For all analyses pertaining to the fingers we considered
only data from the MOS instruments.  The PN data were not used since
chip gaps, bad columns, and the out of time event correction would have
complicated the analysis.  We then compared the average surface
brightness of these regions to that of other regions between the
fingers at the same distance form the center of NGC~4649.  For the
regions containing the fingers, we found on average (1.74$\pm$0.04)$\times
10^{-7}$ counts s$^{-1}$ pix$^{-1}$ and (1.47$\pm$0.03)$\times
10^{-7}$ counts s$^{-1}$ pix$^{-1}$ for the regions without fingers
(1-$\sigma$ errors).  The excess of counts in the regions containing
fingers was therefore significant at the 5.4 $\sigma$ level.

\begin{figure}
\plotone{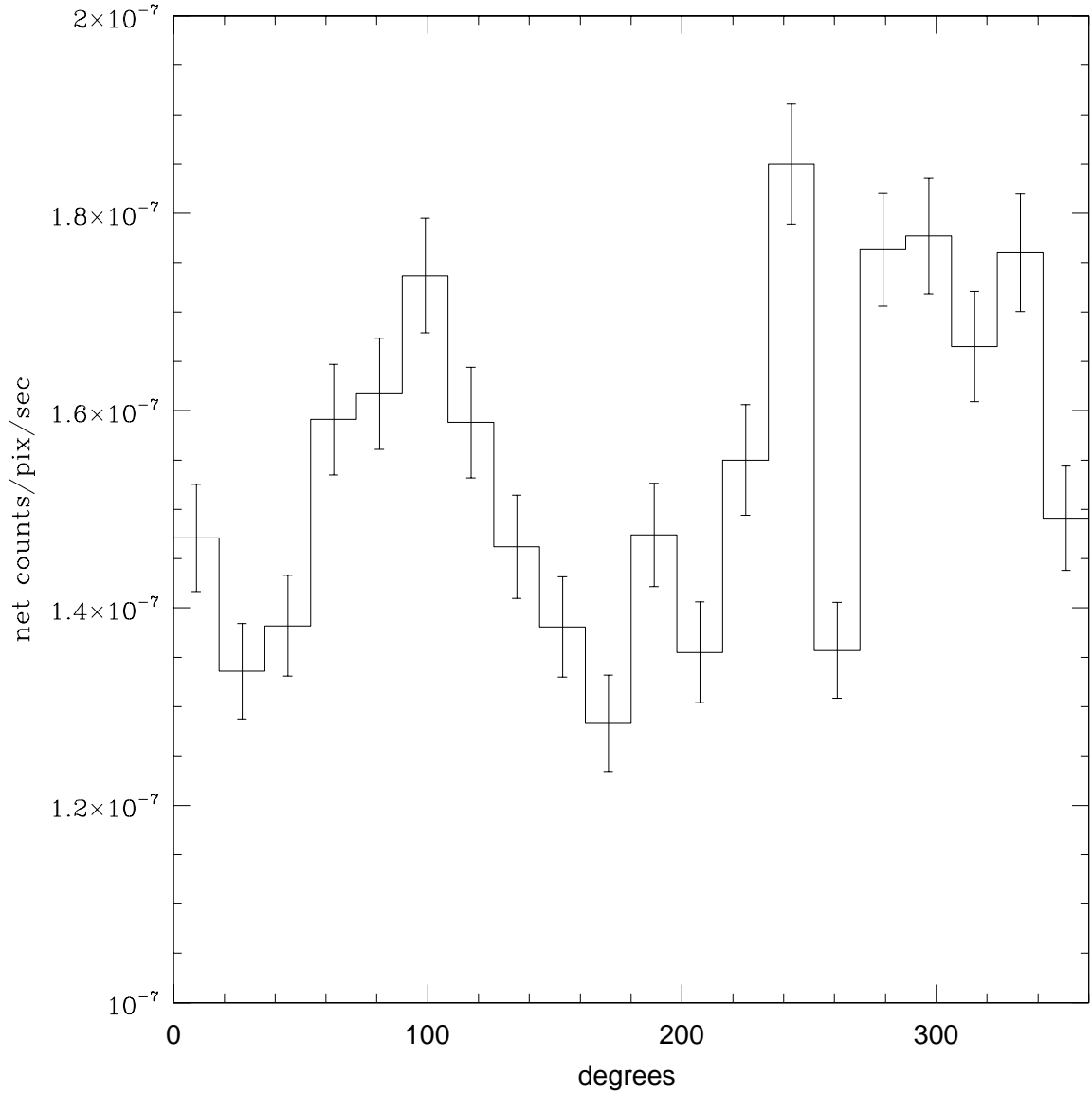}
\caption{Azimuthal plot of the combined net flux of the MOS
instruments with 1-$\sigma$ error bars in
20 angular bins between 25\arcsec\
and 53\arcsec\ from the center of NGC4649.  These counts are
background subtracted but not corrected for exposure.  Angles are
measured from north to east.
\label{fig:azimuth_fingers}}
\end{figure}

Azimuthal brightness variations in the regions of the fingers had
previously been detected in the {\it Chandra} data.
To test for
these variations in the {\it XMM-Newton} data we again divided up an annulus,
ranging from 25\arcsec\ to 53\arcsec\ (2.0-4.3 kpc) and
centered on NGC~4649, into 20 angular bins, excluding {\it Chandra} source
regions.
The data from the two MOS instruments were combined to
determine the total net flux per bin.  An azimuthal plot of the net
flux in each bin
with 1-$\sigma$ errors is given in Figure~\ref{fig:azimuth_fingers}.
Qualitatively this figure is similar to the analogous figure for {\it
Chandra}, though the fingers are somewhat broader, most likely due to
the increased PSF of {\it XMM-Newton}.
Additionally, the fingers are more significantly detected with {\it XMM-Newton}
despite the increased PSF since the larger collecting area and greater
total counts provides a better SNR.
A $\chi^2$ significance test was performed comparing the net flux in
each bin to the mean net flux per bin.  We found a reduced $\chi^2$ of
$\chi^2_{\nu} = 10$, which confirms our previous conclusion that the
azimuthal brightness profile is not well-described by a constant and
that the fingers are likely real features in the diffuse emission.

In our {\it Chandra} analysis we compared the properties of the
observed fingers to those produced in numerical hydrodynamical
simulations of cooling flows in elliptical galaxies by Kritsuk,
B\"ohringer, \& M\"uller (1998).  We now compare these fingers to
those seen in more recent simulations by Kritsuk, Plewa, \& M\"uller
(2001, hereafter KPM), which implement a hybrid model that combines a subsonic
peripheral cooling flow with an inner convective core.
Their simulations
predict outflowing bubbles of hot gas, surrounded by
inflowing gas that is significantly cooler than the
outflowing gas at the same distance from the galactic center, and a
temperature profile that is raised in the center, drops to a minimum
at a few kiloparsecs, then rises again in the outer regions (see their
Figure~7).  The shape of their predicted temperature profile
qualitatively matches what we observe (see
Table~\ref{tab:deproj_spec_grad}), although
KPM find that the temperature profile varies by a factor of two or
more, whereas the variations we observe are on the order of 14\%.
To compare the temperatures of the inflowing and outflowing gas to
those predicted by KPM we fit spectra accumulated from the sum of
the finger regions
and from the regions in between the fingers, using the same MEKAL plus
power-law model used to fit the spectrum of the diffuse emission.
As with {\it Chandra}, the abundances in these fits were fixed at 0.6
solar.  If the
abundances were allowed to vary the resulting values
were unreasonably high and had large errors.
The temperatures and $\chi^2$'s given by the fits were
not significantly affected by fixing the abundance values.
The results of these fits are given in Table~\ref{tab:spectra}.
The best-fit temperatures for the finger and off-finger regions
closely matched those from fitting {\it Chandra} data, with the finger
regions being slightly hotter than the off-finger regions but
consistent within the 90\% confidence intervals
(difference is significant at the 2.1-$\sigma$ level).
In the KPM
model, not only would would expect a larger temperature difference
between the regions, but one would also expect the bright regions to
be cooler than the darker regions, which is the opposite of what we
find.
Although we might expect some scaling differences between our
observations and
the KPM simulations,
we would expect at least the sign of the temperature
difference between the finger and off-finger regions to be the same.
We therefore conclude that while some
form of convective instability may explain the fingers seen in
NGC~4649, the details are not reproduced by the KPM model.

\subsection{Central Bar} \label{sec:bar}

We previously detected a 5\arcsec\ (0.41 kpc) long bar located at the central peak
of the diffuse emission using {\it Chandra} observations of NGC~4649.
Unfortunately, this structure is too small to be resolved by {\it
XMM-Newton}.  However, as discussed in \S~\ref{sec:radial}, {\it XMM-Newton's} large
collecting area provided us with enough counts to deproject the
spectrum of the diffuse emission and obtain more accurate spectral
models for the central emission.  We used the best fit MEKAL plus
power-law model for the central region defined in
Table~\ref{tab:deproj_spec_grad}, applied the appropriate {\it Chandra} response
files, and renormalized the model until the number of net counts
matched that detected for the X-ray bar by {\it Chandra} (about 900
net counts).  The resulting MEKAL component normalization was then
used to calculate the cooling time of the X-ray bar.  We assumed a
prolate ellipsoid geometry for the bar with semi-axes of 1.35\arcsec,
1.35\arcsec, and 3.55\arcsec\ (0.11 kpc by 0.11 kpc by 0.29 kpc).
 The resulting integrated cooling time
was about $t_{\rm cool} \approx 10^6$ yr.  We note that there is
precedence for cooling times of $< 10^7$ yr in the central regions of
elliptical galaxies (e.g., Bregman \& Athey 2004\footnote{Available at
\url{http://www.astro.virginia.edu/coolflow/proceedings/150.pdf}.};
Lowenstein et al. 2001).
This short cooling time suggests that the bar is a transient feature.
One possible cause is accretion onto
the central source detected by Soldatenkov, Vikhlinin, \& Pavlinsky
(2003), which they interpret as a quiescent supermassive black hole.

\section{Filament} \label{sec:filament}

Figure~\ref{fig:xray_whole} appears to show a
faint filament extending from the northeast of NGC~4649 that curves to
the east and extends from about 2\farcm3 to 7\farcm0.  To
determine the statistical significance of this feature we compared the
net counts (0.3--5.0 keV) in a region containing the filament to those
in identical rotated
regions on either side of the filament, with source regions removed.
We found an excess of counts in the filament region at a very high level of
significance, suggesting that it may be a real feature.

\begin{figure}
\centering
\includegraphics[angle=-90,width=6.5in]{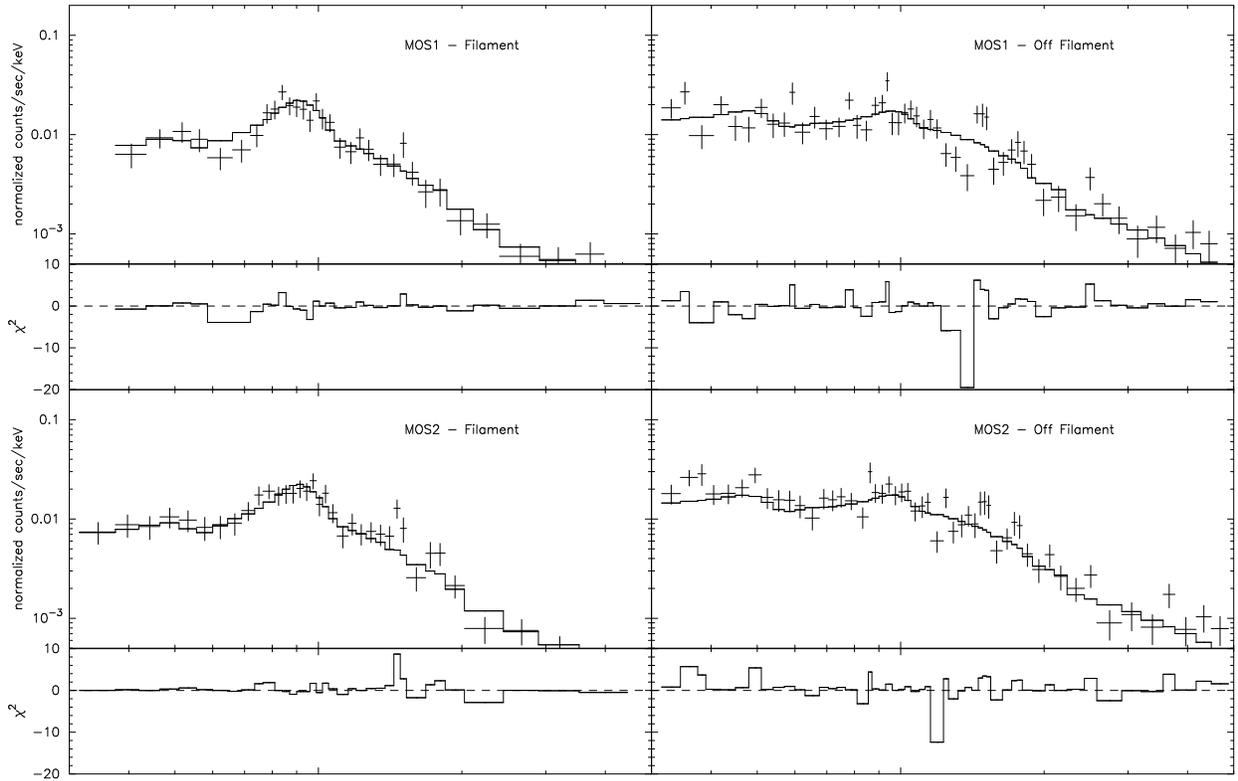}
\caption{
X-ray spectra of the faint filament seen extending to the NE of
NGC~4649 and of two identical regions on either side of the filament,
with source regions removed.
The blend of Fe lines near 0.9 keV is seen to be stronger in the
region of the filament.  The best-fit MEKAL plus powerlaw model is
also shown, where
the abundance has been set to 60\% solar.
\label{fig:filament}}
\end{figure}

Spectra of the filament and off-filament regions are shown in
Figure~\ref{fig:filament}.  Since the filament spectra showed some
emission lines we attempted to model the filament and off-filament
spectra with an absorbed MEKAL model in the 0.3--5.0 keV range
(with absorption fixed at the Galactic value).  This provided a
reasonable fit, with a lower temperature and higher abundance in the
filament region as compared to the off-filament regions (see
Table~\ref{tab:spectra}).  However, the reduced $\chi^2$'s were high.
An examination of the residuals showed that the model underestimated
the hard emission for both the filament and off-filament regions.
Since
we expect some contamination from point sources
in the region (potentially both galactic and background sources), we
added a power-law component to the model, which
provided a much improved fit.  Unfortunately, the resulting abundance
values were unreasonably large and were essentially unconstrained.  We
therefore froze the abundance values at 60\% solar, a value which we
believe to be representative of the abundance in the outer regions of
NGC~4649 (see Table~\ref{tab:proj_spec_grad}).  This fitted model is
shown in Figure~\ref{fig:filament} for the filament and off-filament
regions.  The best-fit
temperatures and their errors were essentially unaffected by freezing
the abundance values.  We conclude that the diffuse emitting gas in the
region of the filament has a lower temperature than the surrounding
gas and may also have a higher abundance (though the latter is unclear).

One potential worry with the above analysis is that the off-axis PSF
is large enough for a significant fraction of source counts to be outside the
18\arcsec\ radius source regions used to remove sources.  Since the
filament contains more detected sources than the neighboring
off-filament regions one might worry that the excess of counts on the
filament was due solely to scattered source counts, though the clear
presence of lines in the filament spectra as compared to the
off-filament spectra would seem to argue against this interpretation.
As a further test, we compared the fluxes from the power-law components
of the filament and off-filament models and found that the flux from
the off-filament component was larger by more than a factor of two.
This suggests that the excess of counts in the region of the filament
is not due mainly to sources, but to diffuse emission.  We conclude
that the filament is a real feature, though further observations would
be useful to verify this claim.

\section{Conclusions} \label{sec:conclusion}

We have presented results from a {\it XMM-Newton} observation of the X-ray
bright elliptical galaxy NGC~4649.  Both bright diffuse emission and
point sources are detected, with the bright diffuse emission dominating the
overall emission.

A total of 158 discrete sources were detected, although we have
limited our discussion to the 47 sources found within four effective
radii of the center of NGC~4649.
Seven of these sources had possible faint
optical counterparts in the DSS image of the same field.
Of these seven, two had positions listed in the USNO-B1.0 catalog.
We find that 23 of the sources match the positions of one or more
sources detected with {\it Chandra}.   Out of the 8 of these
sources with well-determined luminosities 3 showed evidence for a
varying luminosity between the {\it Chandra} and {\it XMM-Newton}
observations.  In addition, {\it XMM-Newton} detected 5 sources not found
with {\it Chandra}, suggesting that these sources brightened
significantly between observations.
Some sources may be associated with the companion Sc galaxy
NGC~4647.  In particular, Src. 21 was within 3\arcsec\ of the optical
center of NGC~4647 and had very soft X-ray colors, which is consistent
with this source being a detection of diffuse gas associated with
NGC~4647.
We did not detect X-rays from the Type I supernova
SN1979a in NGC~4647.

The composite X-ray spectrum of the resolved sources within 4 $R_{\rm
eff}$ is best described by a power-law with a photon spectral index of
$\Gamma \approx 1.76$.
The spectrum of the diffuse emission is reasonably well-described by a
VMEKAL plus power-law model, where the elements are split into three
abundance groups.
This argues that the diffuse emission is
a combination of emission from diffuse gas and unresolved LMXBs and
that the elemental abundances
of the combined diffuse emission
do not
simply scale from solar ratios.
We find a Si/Fe ratio of > 2, suggesting that the diffuse gas was
mainly enriched by Type II SN.
Deprojection of the diffuse emission
shows a temperature profile that is hot in the outer regions of the
galaxy, cools to a minimum temperature closer to the center, and rises
again in the central 0.8--1.6 kpc.  The deprojected spectrum of the central region
is well described by a two-temperature VMEKAL model with abundance
ratios that do not scale from solar, although the abundance values are
poorly determined.

We confirm the existence of faint radial features in the diffuse
emission, seen in {\it Chandra} images (Paper I), extending from the
center of NGC~4649.  The properties of these fingers were compared to
those of similar features seen in simulations using a hybrid cooling
flow/convective core model for giant elliptical galaxies performed by
Kritsuk et al.\ (2001).  We
find that although some properties, such as the galactic
radial temperature profile and the morphology of the fingers, match those
predicted by the simulations, the relative temperatures of the finger
and off-finger regions are inconsistent with the simulations.  We
conclude that although some form of convective instability may explain
these features, they are inconsistent with the specific predictions of
the KPM model.

The deprojected spectrum for the central region was used to model the
diffuse emission in the region of the X-ray bar observed with {\it
Chandra} (Paper I).  We find an integrated cooling time of $t_{\rm
cool} \approx 10^6$ yr in the region of the bar.
We conclude that the bar is a transient hydrodynamical feature.
One possible cause is the central source detected by Soldatenkov, Vikhlinin,
\& Pavlinsky (2003), which they interpret as a quiescent supermassive
black hole.

We find evidence for a filament of diffuse emission, extending to the
NE of NGC~4649 and curving slightly eastward.  The filament appears
to be cooler than the surrounding gas and it may also have a higher
abundance.
It is possible that the filament is the result of a
tidal interaction, possibly with the nearby spiral NGC~4647, although
the symmetric morphology of both NGC~4647 and NGC~4649 would seem to
argue against this interpretation.

\acknowledgements
We are grateful to Arunav Kundu for providing us with his
unpublished list of globular clusters in NGC~4649,
and to the referee for useful suggestions.
This work was supported by NASA {\it XMM-Newton} Grants NAG5-10074
and NAG5-13645.
S. W. R. was supported in part by a fellowship from the Virginia Space
Grant Consortium.
The {\it XMM-Newton} project is an ESA Science Mission with instruments and
contributions directly funded by ESA Member States and the USA (NASA).
This publication makes use of data products from the Two Micron All
Sky Survey, which is a joint project of the University of
Massachusetts and the Infrared Processing and Analysis
Center/California Institute of Technology, funded by the National
Aeronautics and Space Administration and the National Science
Foundation.


\clearpage
\begin{deluxetable}{lcccrrrrccccccl}
\tablewidth{0pt}
\tabletypesize{\tiny}
\tablecolumns{15}
\tablecaption{Discrete X-ray Sources \label{tab:src}}
\rotate
\tablehead{
\colhead{Src.}&
\colhead{Name}&
\colhead{R.A.}&
\colhead{Dec.}&
\colhead{$d$}&
\colhead{Count Rate}&
\colhead{SNR}&
\colhead{$L_X$}&
\colhead{H21}&
\colhead{H31}&
\colhead{PN}&
\colhead{M1}&
\colhead{M2}&
\colhead{}
\\
\colhead{No.}&
\colhead{}&
\colhead{(h:m:s)}&
\colhead{($\arcdeg$:$\arcmin$:$\arcsec$)}&
\colhead{($\arcsec$)}&
\colhead{($10^{-4}$ s$^{-1}$)}&
\colhead{}&
\colhead{}&
\colhead{}&
\colhead{}&
\colhead{}&
\colhead{}&
\colhead{}&
\colhead{Notes} \\
\colhead{(1)}&
\colhead{(2)}&
\colhead{(3)}&
\colhead{(4)}&
\colhead{(5)}&
\colhead{(6)}&
\colhead{(7)}&
\colhead{(8)}&
\colhead{(9)}&
\colhead{(10)}&
\colhead{(11)}&
\colhead{(12)}&
\colhead{(13)}&
\colhead{(14)}
}
\startdata
1&XMMJJ 124340.2+113308&12:43:40.18&11:33:08.4&2.94&12417.78$\pm$\phn77.97&161.94&1646.73&$-0.39^{+0.01}_{-0.01}$&$-0.88^{+0.00}_{-0.00}$&SMHT&SMHT&SMHT&a\\
2&XMMJJ 124338.0+113210&12:43:37.96&11:32:10.7&66.76&37.90$\pm$\phn\phn7.67&4.97&5.03&$-0.86^{+0.17}_{-0.08}$&$-0.91^{+0.13}_{-0.05}$&S\phm{M}\phm{H}\phm{T}&S\phm{M}\phm{H}T&\phm{S}\phm{M}\phm{H}\phm{T}&b\\
3&XMMJJ 124336.4+113220&12:43:36.43&11:32:20.5&72.45&22.25$\pm$\phn\phn6.59&3.42&2.95&$+0.29^{+0.23}_{-0.26}$&$-0.41^{+0.16}_{-0.14}$&\phm{S}M\phm{H}\phm{T}&\phm{S}\phm{M}\phm{H}\phm{T}&\phm{S}\phm{M}\phm{H}\phm{T}&b\\
4&XMMJJ 124345.1+113233&12:43:45.13&11:32:33.1&83.74&48.12$\pm$\phn\phn5.29&9.12&6.38&$-0.15^{+0.11}_{-0.10}$&$-0.45^{+0.18}_{-0.13}$&\phm{S}\phm{M}HT&\phm{S}MHT&\phm{S}\phm{M}HT&c,d\\
5&XMMJJ 124334.9+113228&12:43:34.94&11:32:28.1&85.72&105.25$\pm$\phn\phn8.95&11.79&13.96&$+0.30^{+0.10}_{-0.11}$&$-0.08^{+0.13}_{-0.13}$&SMHT&\phm{S}MHT&SM\phm{H}T&b,c,e\\
6&XMMJJ 124334.6+113343&12:43:34.56&11:33:43.2&86.77&65.07$\pm$\phn\phn7.71&8.52&8.63&$+0.58^{+0.10}_{-0.12}$&$+0.24^{+0.16}_{-0.18}$&\phm{S}MHT&\phm{S}M\phm{H}T&S\phm{M}\phm{H}T&\\
7&XMMJJ 124337.4+113143&12:43:37.45&11:31:43.5&94.60&297.50$\pm$\phn12.01&24.83&39.45&$-0.41^{+0.04}_{-0.04}$&$-0.55^{+0.04}_{-0.04}$&SMHT&SMHT&SMHT&b,c\\
8&XMMJJ 124334.4+113358&12:43:34.39&11:33:58.2&95.70&43.50$\pm$\phn\phn6.85&6.45&5.77&$+0.49^{+0.18}_{-0.23}$&$+0.30^{+0.23}_{-0.27}$&\phm{S}\phm{M}\phm{H}T&\phm{S}M\phm{H}T&\phm{S}\phm{M}HT&\\
9&XMMJJ 124346.3+113336&12:43:46.31&11:33:36.0&95.97&11.49$\pm$\phn\phn3.54&3.23&1.52&$-0.10^{+0.10}_{-0.08}$&$-0.39^{+0.12}_{-0.11}$&\phm{S}\phm{M}\phm{H}\phm{T}&\phm{S}\phm{M}\phm{H}\phm{T}&\phm{S}\phm{M}\phm{H}\phm{T}&c\\
10&XMMJJ 124342.7+113438&12:43:42.73&11:34:38.3&96.69&16.95$\pm$\phn\phn4.44&3.83&2.25&$-0.19^{+0.18}_{-0.11}$&$-0.55^{+0.13}_{-0.10}$&\phm{S}\phm{M}\phm{H}\phm{T}&\phm{S}\phm{M}\phm{H}\phm{T}&\phm{S}\phm{M}H\phm{T}&c,d\\
11&XMMJJ 124335.8+113155&12:43:35.76&11:31:55.7&97.35&54.03$\pm$\phn\phn7.56&7.23&7.17&$-0.24^{+0.14}_{-0.13}$&$-0.79^{+0.16}_{-0.10}$&\phm{S}M\phm{H}T&S\phm{M}\phm{H}T&S\phm{M}\phm{H}T&\\
12&XMMJJ 124335.8+113425&12:43:35.77&11:34:26.0&98.22&79.48$\pm$\phn\phn7.69&10.50&10.54&$-0.01^{+0.11}_{-0.11}$&$-0.18^{+0.12}_{-0.12}$&SMHT&\phm{S}M\phm{H}T&SM\phm{H}T&c,d,e,f,g\\
13&XMMJJ 124333.4+113240&12:43:33.41&11:32:40.6&101.48&75.26$\pm$\phn\phn7.34&10.38&9.98&$+0.43^{+0.14}_{-0.17}$&$+0.40^{+0.15}_{-0.17}$&\phm{S}MHT&\phm{S}MHT&\phm{S}MHT&b,c\\
14&XMMJJ 124347.3+113235&12:43:47.29&11:32:35.4&112.32&119.64$\pm$\phn\phn9.22&13.11&15.87&$+0.41^{+0.14}_{-0.17}$&$+0.69^{+0.08}_{-0.10}$&\phm{S}MHT&\phm{S}MHT&\phm{S}\phm{M}HT&c\\
15&XMMJJ 124335.9+113123&12:43:35.92&11:31:23.1&122.88&20.24$\pm$\phn\phn5.77&3.50&2.68&$+0.24^{+0.32}_{-0.39}$&$+0.09^{+0.36}_{-0.39}$&\phm{S}\phm{M}\phm{H}\phm{T}&\phm{S}\phm{M}\phm{H}\phm{T}&\phm{S}\phm{M}\phm{H}\phm{T}&c,h\\
16&XMMJJ 124348.7+113302&12:43:48.72&11:33:03.0&128.10&8.81$\pm$\phn\phn2.84&3.06&1.17&$+0.29^{+0.22}_{-0.26}$&$+0.28^{+0.22}_{-0.25}$&\phm{S}\phm{M}\phm{H}\phm{T}&\phm{S}\phm{M}\phm{H}\phm{T}&\phm{S}\phm{M}\phm{H}\phm{T}&c\\
17&XMMJJ 124349.0+113242&12:43:49.00&11:32:42.7&134.78&22.50$\pm$\phn\phn5.20&4.39&2.98&$+0.05^{+0.37}_{-0.39}$&$+0.46^{+0.24}_{-0.33}$&\phm{S}\phm{M}H\phm{T}&\phm{S}\phm{M}\phm{H}\phm{T}&\phm{S}\phm{M}\phm{H}\phm{T}&h\\
18&XMMJJ 124332.1+113418&12:43:32.06&11:34:18.1&135.24&19.85$\pm$\phn\phn4.05&4.87&2.63&$+0.67^{+0.23}_{-0.50}$&$+0.82^{+0.13}_{-0.33}$&\phm{S}\phm{M}\phm{H}T&\phm{S}\phm{M}\phm{H}\phm{T}&\phm{S}\phm{M}H\phm{T}&b,c,g\\
19&XMMJJ 124332.2+113143&12:43:32.24&11:31:43.3&143.68&37.19$\pm$\phn\phn5.16&7.30&4.93&$-0.23^{+0.38}_{-0.32}$&$+0.63^{+0.14}_{-0.20}$&\phm{S}\phm{M}HT&\phm{S}\phm{M}HT&\phm{S}\phm{M}HT&c\\
20&XMMJJ 124330.1+113318&12:43:30.11&11:33:18.5&145.85&22.79$\pm$\phn\phn3.73&6.06&6.96&$+0.32^{+0.25}_{-0.30}$&$+0.62^{+0.14}_{-0.20}$&\phm{S}\phm{M}\phm{H}\phm{T}&\phm{S}\phm{M}HT&\phm{S}\phm{M}HT&b,c,i\\
21&XMMJJ 124332.7+113453&12:43:32.71&11:34:53.2&148.85&77.43$\pm$\phn\phn6.82&11.39&10.27&$-0.83^{+0.08}_{-0.06}$&$-0.92^{+0.06}_{-0.03}$&S\phm{M}\phm{H}T&S\phm{M}\phm{H}T&S\phm{M}\phm{H}T&b,g,j,k\\
22&XMMJJ 124347.6+113529&12:43:47.65&11:35:29.9&179.14&18.38$\pm$\phn\phn4.77&3.95&2.44&$+0.03^{+0.26}_{-0.26}$&$+0.02^{+0.27}_{-0.27}$&\phm{S}\phm{M}\phm{H}\phm{T}&\phm{S}\phm{M}\phm{H}T&\phm{S}\phm{M}\phm{H}\phm{T}&\\
23&XMMJJ 124351.8+113356&12:43:51.81&11:33:56.0&179.19&9.63$\pm$\phn\phn3.08&3.16&1.28&$+0.58^{+0.20}_{-0.32}$&$+0.56^{+0.21}_{-0.32}$&\phm{S}\phm{M}\phm{H}\phm{T}&\phm{S}\phm{M}\phm{H}\phm{T}&\phm{S}\phm{M}\phm{H}\phm{T}&b,c\\
24&XMMJJ 124341.3+113009&12:43:41.28&11:30:09.7&181.40&13.91$\pm$\phn\phn3.97&3.51&1.85&$-0.07^{+0.42}_{-0.40}$&$+0.28^{+0.28}_{-0.34}$&\phm{S}\phm{M}\phm{H}\phm{T}&\phm{S}\phm{M}\phm{H}\phm{T}&\phm{S}\phm{M}\phm{H}\phm{T}&c\\
25&XMMJJ 124348.9+113518&12:43:48.86&11:35:18.2&182.39&13.48$\pm$\phn\phn4.25&3.20&1.79&$-0.35^{+0.56}_{-0.39}$&$+0.25^{+0.31}_{-0.37}$&\phm{S}\phm{M}\phm{H}\phm{T}&\phm{S}\phm{M}\phm{H}\phm{T}&\phm{S}\phm{M}\phm{H}\phm{T}&\\
26&XMMJJ 124328.0+113402&12:43:28.03&11:34:02.5&183.75&20.24$\pm$\phn\phn3.94&5.18&2.68&$+0.56^{+0.21}_{-0.32}$&$+0.69^{+0.15}_{-0.25}$&\phm{S}M\phm{H}T&\phm{S}\phm{M}\phm{H}\phm{T}&\phm{S}\phm{M}HT&b,c\\
27&XMMJJ 124335.9+113607&12:43:35.88&11:36:07.0&186.99&23.87$\pm$\phn\phn4.16&5.90&3.17&$+0.04^{+0.21}_{-0.22}$&$+0.20^{+0.18}_{-0.19}$&\phm{S}\phm{M}\phm{H}\phm{T}&S\phm{M}\phm{H}T&\phm{S}\phm{M}HT&b\\
28&XMMJJ 124343.1+113008&12:43:43.07&11:30:08.4&187.24&39.07$\pm$\phn\phn4.95&7.89&5.18&$-0.21^{+0.13}_{-0.13}$&$-0.49^{+0.17}_{-0.14}$&SM\phm{H}T&S\phm{M}\phm{H}T&\phm{S}\phm{M}\phm{H}T&c\\
29&XMMJJ 124336.6+113008&12:43:36.65&11:30:08.5&188.38&665.85$\pm$\phn14.84&44.95&88.30&$-0.30^{+0.02}_{-0.02}$&$-0.50^{+0.02}_{-0.02}$&SMHT&SMHT&SMHT&b,c,e,h,j\\
30&XMMJJ 124344.6+113003&12:43:44.57&11:30:03.5&198.35&25.76$\pm$\phn\phn4.06&6.40&3.42&$-0.05^{+0.22}_{-0.21}$&$+0.14^{+0.18}_{-0.19}$&\phm{S}\phm{M}\phm{H}T&\phm{S}\phm{M}HT&\phm{S}\phm{M}\phm{H}T&b,c\\
31&XMMJJ 124347.1+113020&12:43:47.09&11:30:20.7&198.82&16.24$\pm$\phn\phn3.80&4.33&2.15&$+0.16^{+0.27}_{-0.30}$&$+0.18^{+0.26}_{-0.29}$&\phm{S}\phm{M}\phm{H}\phm{T}&\phm{S}\phm{M}\phm{H}T&\phm{S}\phm{M}\phm{H}\phm{T}&b,c\\
32&XMMJJ 124346.6+113008&12:43:46.59&11:30:08.1&206.15&16.09$\pm$\phn\phn3.75&4.28&2.13&$+0.28^{+0.29}_{-0.35}$&$+0.45^{+0.23}_{-0.31}$&\phm{S}\phm{M}\phm{H}T&\phm{S}\phm{M}\phm{H}\phm{T}&\phm{S}\phm{M}\phm{H}\phm{T}&b\\
33&XMMJJ 124332.8+113001&12:43:32.84&11:30:01.3&216.38&116.04$\pm$\phn\phn6.55&17.82&15.39&$-0.18^{+0.06}_{-0.05}$&$-0.63^{+0.07}_{-0.06}$&SMHT&SMHT&SM\phm{H}T&\\
34&XMMJJ 124337.3+113646&12:43:37.26&11:36:46.7&220.23&27.64$\pm$\phn\phn3.61&7.83&3.67&$+0.03^{+0.13}_{-0.13}$&$-0.17^{+0.17}_{-0.16}$&\phm{S}\phm{M}\phm{H}\phm{T}&SMHT&SM\phm{H}T&b,c\\
35&XMMJJ 124330.9+113008&12:43:30.93&11:30:08.6&225.48&26.72$\pm$\phn\phn4.25&6.24&3.54&$-0.16^{+0.19}_{-0.18}$&$-0.16^{+0.19}_{-0.18}$&SM\phm{H}T&\phm{S}\phm{M}\phm{H}\phm{T}&\phm{S}\phm{M}\phm{H}\phm{T}&\\
36&XMMJJ 124348.6+113619&12:43:48.56&11:36:19.2&226.83&11.44$\pm$\phn\phn3.60&3.25&1.52&$-0.32^{+0.34}_{-0.27}$&$-0.36^{+0.33}_{-0.26}$&\phm{S}\phm{M}\phm{H}\phm{T}&\phm{S}\phm{M}\phm{H}\phm{T}&\phm{S}\phm{M}\phm{H}\phm{T}&\\
37&XMMJJ 124326.9+113515&12:43:26.94&11:35:15.2&229.24&32.73$\pm$\phn\phn4.22&7.82&4.34&$+0.57^{+0.13}_{-0.17}$&$+0.33^{+0.20}_{-0.23}$&\phm{S}M\phm{H}T&\phm{S}\phm{M}\phm{H}T&\phm{S}M\phm{H}T&c,h,g\\
38&XMMJJ 124325.8+113120&12:43:25.85&11:31:20.6&235.37&58.55$\pm$\phn\phn4.75&12.33&7.76&$+0.09^{+0.10}_{-0.10}$&$-0.02^{+0.11}_{-0.11}$&SMHT&SMHT&SMHT&b,h\\
39&XMMJJ 124342.1+113727&12:43:42.06&11:37:28.0&259.50&23.24$\pm$\phn\phn3.65&6.46&3.08&$+0.17^{+0.18}_{-0.19}$&$+0.09^{+0.20}_{-0.21}$&\phm{S}\phm{M}\phm{H}T&\phm{S}M\phm{H}T&\phm{S}\phm{M}\phm{H}T&c,h\\
40&XMMJJ 124353.3+113618&12:43:53.32&11:36:18.2&271.22&12.29$\pm$\phn\phn3.28&3.68&1.63&$-0.58^{+0.30}_{-0.19}$&$-0.90^{+0.38}_{-0.08}$&S\phm{M}\phm{H}T&\phm{S}\phm{M}\phm{H}\phm{T}&\phm{S}\phm{M}\phm{H}\phm{T}&\\
\enddata
\end{deluxetable}

\clearpage
\begin{deluxetable}{lcccrrrrccccccl}
\tablewidth{0pt}
\tabletypesize{\tiny}
\tablecolumns{15}
\rotate
\tablenum{1}
\tablecaption{---     $Continued$}
\tablehead{
\colhead{Src.}&
\colhead{Name}&
\colhead{R.A.}&
\colhead{Dec.}&
\colhead{$d$}&
\colhead{Count Rate}&
\colhead{SNR}&
\colhead{$L_X$}&
\colhead{H21}&
\colhead{H31}&
\colhead{PN}&
\colhead{M1}&
\colhead{M2}&
\colhead{}
\\
\colhead{No.}&
\colhead{}&
\colhead{(h:m:s)}&
\colhead{($\arcdeg$:$\arcmin$:$\arcsec$)}&
\colhead{($\arcsec$)}&
\colhead{($10^{-4}$ s$^{-1}$)}&
\colhead{}&
\colhead{}&
\colhead{}&
\colhead{}&
\colhead{}&
\colhead{}&
\colhead{}&
\colhead{Notes} \\
\colhead{(1)}&
\colhead{(2)}&
\colhead{(3)}&
\colhead{(4)}&
\colhead{(5)}&
\colhead{(6)}&
\colhead{(7)}&
\colhead{(8)}&
\colhead{(9)}&
\colhead{(10)}&
\colhead{(11)}&
\colhead{(12)}&
\colhead{(13)}&
\colhead{(14)}
}
\startdata
41&XMMJJ 124325.6+113017&12:43:25.62&11:30:18.0&272.83&23.70$\pm$\phn\phn3.07&7.64&7.24&$+0.20^{+0.16}_{-0.17}$&$+0.33^{+0.14}_{-0.16}$&\phm{S}\phm{M}\phm{H}\phm{T}&SMHT&\phm{S}M\phm{H}T&b,h,i,j\\
42&XMMJJ 124356.1+113051&12:43:56.06&11:30:51.7&273.48&12.07$\pm$\phn\phn3.29&3.63&1.60&$-0.18^{+0.33}_{-0.30}$&$-0.08^{+0.32}_{-0.31}$&\phm{S}\phm{M}\phm{H}T&\phm{S}\phm{M}\phm{H}\phm{T}&\phm{S}\phm{M}\phm{H}\phm{T}&b\\
43&XMMJJ 124327.4+112938&12:43:27.40&11:29:38.2&281.61&47.83$\pm$\phn\phn4.29&11.14&6.34&$-0.20^{+0.11}_{-0.10}$&$-0.09^{+0.11}_{-0.10}$&SMHT&SMHT&\phm{S}\phm{M}HT&\\
44&XMMJJ 124320.8+113240&12:43:20.79&11:32:40.8&284.09&17.51$\pm$\phn\phn3.47&5.00&2.32&$-0.26^{+0.21}_{-0.19}$&$-0.42^{+0.27}_{-0.21}$&S\phm{M}\phm{H}T&\phm{S}\phm{M}\phm{H}\phm{T}&\phm{S}\phm{M}\phm{H}\phm{T}&\\
45&XMMJJ 124320.4+113504&12:43:20.43&11:35:04.3&309.64&27.20$\pm$\phn\phn3.74&7.36&3.61&$+0.30^{+0.22}_{-0.25}$&$+0.61^{+0.13}_{-0.18}$&\phm{S}\phm{M}HT&\phm{S}\phm{M}HT&\phm{S}\phm{M}HT&\\
46&XMMJJ 124337.5+112801&12:43:37.48&11:28:01.4&311.03&45.23$\pm$\phn\phn4.50&10.10&6.00&$+0.15^{+0.12}_{-0.13}$&$+0.19^{+0.12}_{-0.13}$&SMHT&\phm{S}MHT&\phm{S}\phm{M}HT&\\
47&XMMJJ 124357.4+113625&12:43:57.37&11:36:25.8&321.34&15.34$\pm$\phn\phn3.02&5.11&2.03&$+0.17^{+0.23}_{-0.25}$&$+0.17^{+0.24}_{-0.27}$&\phm{S}\phm{M}\phm{H}\phm{T}&\phm{S}M\phm{H}T&\phm{S}\phm{M}\phm{H}T&b,j\\
\enddata
\tablecomments{The units for $L_X$ are $10^{38}$ ergs s$^{-1}$ in the
0.3--12 keV band.}
\tablenotetext{a}{\scriptsize Src.~1 is is extended, and appears to be a
combination of diffuse structure with one or more point sources.}
\tablenotetext{b}{\scriptsize Source is near a chip gap or bad column
on one or more instruments.  Position and count rate may be inaccurate}
\tablenotetext{c}{\scriptsize Possible Chandra X-ray counterpart.}
\tablenotetext{d}{\scriptsize May be a blend of two Chandra sources.}
\tablenotetext{e}{\scriptsize Globular cluster is possible optical counterpart.}
\tablenotetext{f}{\scriptsize Source may be variable.}
\tablenotetext{g}{\scriptsize May be associated with the companion
galaxy NGC~4647.}
\tablenotetext{h}{\scriptsize Possible faint optical counterpart.}
\tablenotetext{i}{\scriptsize Source was not visible on the PN camera
due to a chip gap or bad column.}
\tablenotetext{j}{\scriptsize Possible USNO-B1.0 optical counterpart.}
\tablenotetext{k}{\scriptsize Possible 2MASS near-infrared counterpart.}
\end{deluxetable}

\clearpage

\begin{deluxetable}{lcccccccc}
\tabletypesize{\small}
\tablewidth{8.5truein}
\tablecaption{Spectral Fits \label{tab:spectra}}
\rotate
\tablehead{
\colhead{Origin}&
\colhead{Model}&
\colhead{$N_H$}&
\colhead{$kT_s$\tablenotemark{a}}&
\colhead{Abund.\tablenotemark{b}}&
\colhead{$\Gamma$ or $k T_h$}&
\colhead{$\chi^2$/dof}&
\colhead{Net Cts.}\\
\colhead{}&
\colhead{}&
\colhead{10$^{20}$cm$^{-2}$}&
\colhead{(keV)}&
\colhead{(solar)}&
\colhead{(keV)}&
\colhead{}&
\colhead{}
}
\startdata
Sources&bremsstrahlung&(2.2)&&&4.40$^{+0.80}_{-0.60}$&653/770=0.85&14464\\
Sources&powerlaw&(2.2)&&&1.76$^{+0.05}_{-0.05}$&575/770=0.75&14464\\
Diffuse\tablenotemark{c}&mekal&(2.2)&0.823$^{+0.004}_{-0.004}$&0.37$^{+0.01}_{-0.01}$&&5522/988=5.59&72587\\
Diffuse\tablenotemark{c}&mekal$+$powerlaw&(2.2)&0.794$^{+0.004}_{-0.004}$&0.81$^{+0.07}_{-0.06}$&(1.78)&1307/427=3.06&72587\\
Diffuse\tablenotemark{c}&mekal$+$mekal$+$powerlaw&(2.2)&0.766$^{+0.009}_{-0.005}$/1.512$^{+0.249}_{-0.112}$&1.08$^{+0.01}_{-0.17}$&(1.78)&1065/425=2.50&72587\\
Diffuse\tablenotemark{c}&vmekal$+$powerlaw&(2.2)&$0.743^{+0.005}_{-0.005}$&$0.22^{+0.08}_{-0.08}$&(1.78)&578/425=1.36&72587\\
&&&&1.67$^{+0.18}_{-0.19}$&&&\\
&&&&$0.70^{+0.06}_{-0.07}$&&&\\
Fingers\tablenotemark{c}&mekal$+$powerlaw&(2.2)&$0.802^{+0.025}_{-0.024}$&(0.6)&(1.78)&119/75=1.59&1978\\
Off Fingers\tablenotemark{c}&mekal$+$powerlaw&(2.2)&$0.761^{+0.020}_{-0.020}$&(0.6)&(1.78)&116/77=1.51&2200\\
Filament\tablenotemark{c}&mekal&(2.2)&$0.929^{+0.070}_{-0.062}$&$0.09^{+}_{-0.02}$&&117/66=1.77&1427\\
Filament\tablenotemark{c}&mekal$+$powerlaw&(2.2)&0.765$^{+0.062}_{-0.053}$&0.17$^{+0.88}_{-0.07}$&1.57$^{+0.57}_{-0.91}$&65/64=1.02&1427\\
Off Filament\tablenotemark{c}&mekal&(2.2)&1.66$^{+0.25}_{-0.21}$&0.02$^{+0.04}_{-0.02}$&&254/101=2.52&2149\\
Off Filament\tablenotemark{c}&mekal$+$powerlaw&(2.2)&0.847$^{+0.20}_{-0.15}$&(0.17)&2.06$^{+0.10}_{-0.10}$&194/100=1.94&2149\\
\enddata
\tablenotetext{a}{For the model with two MEKAL components the
temperature for each component is given.  The abundances are
set equal to one another.}
\tablenotetext{b}{For the VMEKAL model, the first entry in this column
gives the abundance of C, N, O, and Ca; the second of Ne, Mg, Si, S, Ar,
Al, Na, and Ni; and the third simply gives the Fe abundance.}
\tablenotetext{c}{Only data from the MOS instruments were used for
these fits.}
\end{deluxetable}

\clearpage

\begin{deluxetable}{ccccc}
\tablewidth{5.3truein}
\tablecaption{Radial Variation in the Projected Diffuse Spectrum \label{tab:proj_spec_grad}}
\tablehead{
\colhead{Radii}&
\colhead{$kT$}&
\colhead{Abund.}&
\colhead{$\chi^2$/dof}&
\colhead{Net Cts.}\\
\colhead{(\arcsec)}&
\colhead{(keV)}&
\colhead{(solar)}&
\colhead{}&
\colhead{(MOS1/MOS2)}
}
\startdata
0 - 10   &0.820$^{+0.010}_{-0.010}$&2.3$^{+1.4}_{-0.7}$&442/199=2.22&5395/5729\\
10 - 20  &0.781$^{+0.009}_{-0.011}$&1.8$^{+2.3}_{-0.1}$&349/206=1.69&6181/6388\\
20 - 30  &0.746$^{+0.010}_{-0.010}$&2.0$^{+1.7}_{-0.6}$&295/178=1.70&4425/4392\\
30 - 40  &0.765$^{+0.012}_{-0.012}$&8.2$^{+7.0}_{-5.3}$&247/157=1.57&3350/3270\\
40 - 52  &0.764$^{+0.013}_{-0.012}$&2.3$^{+3.6}_{-1.0}$&194/153=1.27&3145/3243\\
52 - 68  &0.809$^{+0.014}_{-0.014}$&1.4$^{+0.9}_{-0.4}$&226/156=1.45&3057/3125\\
68 - 93  &0.814$^{+0.014}_{-0.015}$&2.7$^{+63.7}_{-1.0}$&246/170=1.45&3128/3326\\
93 - 130  &0.838$^{+0.020}_{-0.020}$&0.8$^{+0.5}_{-0.2}$&209/182=1.15&3074/3172\\
130 - 194  &0.848$^{+0.028}_{-0.027}$&0.2$^{+0.05}_{-0.04}$&267/231=1.16&3030/3306\\
\enddata
\end{deluxetable}

\clearpage

\begin{deluxetable}{cccc}
\tablewidth{3.8truein}
\tablecaption{Radial Variation in the Deprojected Diffuse Spectrum\label{tab:deproj_spec_grad}}
\tablehead{
\colhead{Radii}&
\colhead{$kT$}&
\colhead{Abund.}&
\colhead{$\chi^2$/dof}\\
\colhead{(\arcsec)}&
\colhead{(keV)}&
\colhead{(solar)}&
\colhead{}
}
\startdata
0 - 10    &0.845$^{+0.017}_{-0.017}$&2.2$^{+3.1}_{-0.8}$&432/197=2.19\\
10 - 20   &0.803$^{+0.013}_{-0.014}$&2.5$^{+5.3}_{-1.2}$&342/204=1.68\\
20 - 30   &0.724$^{+0.017}_{-0.017}$&1.7$^{+3.2}_{-0.6}$&291/172=1.69\\
30 - 40   &0.773$^{+0.25}_{-0.024}$&29.2$^{+*}_{-*}$&243/155/=1.57\\
40 - 52   &0.717$^{+0.023}_{-0.025}$&5.7$^{+11.2}_{-1.1}$&191/151=1.26\\
52 - 68   &0.804$^{+0.031}_{-0.029}$&1.1$^{+0.3}_{-0.3}$&226/154=1.47\\
68 - 93   &0.802$^{+0.021}_{-0.021}$&13.3$^{+*}_{-*}$&244/168=1.45\\
93 - 130  &0.835$^{+0.029}_{-0.028}$&2.5$^{+10.9}_{-0.7}$&209/180=1.16\\
130 - 194 &0.848$^{+0.028}_{-0.027}$&0.23$^{+0.03}_{-0.02}$&256/229=1.12\\
\enddata
\tablenotetext{*}{These abundance values were undetermined and useful
  confidence intervals could not be found.  See discussion in text.}
\end{deluxetable}


\clearpage

\clearpage

\clearpage

\clearpage

\end{document}